\documentclass[a4paper,11pt]{article}

\usepackage{jinstpub} 
\usepackage{lineno}
\usepackage{hyperref}
\usepackage{siunitx}
\usepackage[caption=false]{subfig}
\usepackage{amsmath}
\usepackage{booktabs}
\usepackage{multirow}
\usepackage{textcomp}
\usepackage{siunitx}

\title{\boldmath Measurements of enriched $^{155}$Gd and $^{157}$Gd converters with the NMX detector on the n\_TOF EAR2 beam line at CERN }

\author[a,b,c]{D. Pfeiffer \note{Corresponding author.}}
\author[b]{F.M. Brunbauer}
\author[b,c]{I.R. Fehse}
\author[a]{A.D. Finke}
\author[a]{K. Fissum}
\author[b]{K.J. Floethner}
\author[b]{D. Janssens}
\author[b]{M. Lisowska}
\author[b,f]{H. Muller}
\author[a]{E. Oksanen}
\author[b]{E. Oliveri}
\author[b]{L. Ropelewski}
\author[d]{A. Rusu}
\author[a,b]{J. Samarati}
\author[b]{L. Scharenberg}
\author[b]{M. van Stenis}
\author[b,e]{R. Veenhof}
\author[b,f]{N. Zavaritskaya}

\affiliation[a]{European Spallation Source ESS ERIC (ESS), \\Box 176, SE-221 00 Lund, Sweden}
\affiliation[b]{European Organization for Nuclear Research (CERN), \\1211 Geneva 23, Switzerland}
\affiliation[c]{Friedrich-Abel-Gymnasium, Alter Postweg 6, 71665 Vaihingen Enz, Germany}
\affiliation[d]{SRS Technology, 30 Promenade des Artisans, 1217 Meyrin, Switzerland}
\affiliation[e]{Bursa Uluda{\u g} University, \\G{\"o}r{\"u}kle Kampusu, 16059 Ni{\"u}fer/Bursa, Turkey}
\affiliation[f]{Lycee International, Saint Genis-Pouilly, France}

\emailAdd{dorothea.pfeiffer@ess.eu}

\abstract{The detectors for the NMX instrument at the European Spallation Source (ESS) in Lund use natural Gd as the neutron converter. In 2024, beam time was obtained at the neutron time-of flight experiment (n\_TOF) at CERN to study the feasibility of an upgrade to enriched Gd. A 10 x 10~cm$^{2}$ prototype of the NMX detector was equipped with two enriched Gd samples ($^{157}$Gd and $^{155}$Gd) that were attached with copper tape to the natural Gd cathode of the detector. Three sets of measurements were taken, with the beam focused on either the natural Gd, the $^{157}$Gd, or the $^{155}$Gd samples. Using the time-of-flight technique with the subsequent conversion of time-of-flight into energy, the resonant region between 1 eV and 200 eV of the $^{157}$Gd and $^{155}$Gd cross sections was studied. The peaks in the resonant region were clearly visible, having higher ADC values in the ADC spectrum. Additionally, the resonant peaks had a larger number of counts per energy bin. In the thermal neutron energy range, the count rate at the center of the beam was measured for natural Gd, $^{157}$Gd, and $^{155}$Gd. Enriched $^{157}$Gd showed an efficiency that was between 60 $\--$ 180 $\%$ higher, compared to natural Gd, for neutron wavelengths between 0.8 \AA~ and 1.8 \AA. The measured 60 $\%$ increase in efficiency at 1.8 \AA~ is lower than expected from mathematical modelling (100 $\%$) and previous measurements with solid state detectors (80 $\%$). Detection of capture gammas and gamma background, bad focusing and saturation effects most likely explain this deviation. An upgrade of the natural Gd converter to enriched $^{157}$Gd would thus lead to an efficiency increase of at least 60 $\%$. The measurements presented in this paper are the first successful time-of-flight measurements with the NMX detector prototype and the ESS VMM readout.}


\begin{document}
\maketitle
\flushbottom

% main text
\section{Introduction}
\label{introduction}
The European Spallation Source (ESS)~\cite{ESS} in Lund, Sweden will become the world's most intense thermal neutron source with a significantly higher brilliance than at existing reactor or spallation sources. For the 15 neutron instruments that will comprise the initial instrument suite at ESS~\cite{Peggs2013, Andersen2020}, efficient thermal neutron detectors are a critical component~\cite{Kirstein2014}. The scientific user program for the first group of five instruments is expected to start in spring 2027. The macromolecular single-crystal diffractometer NMX~\cite{Andersen2020} belongs to this first group, and requires three 51.2 x 51.2~cm$^{2}$ detectors with at least reasonable detection efficiency (around 20$\%$ for cold neutrons in the center of the wavelength range from 1.8 to 10 \AA~), sub-mm spatial resolution~\cite{NMX2020} and a good (~0.1 ms) time resolution. The combination of solid neutron converters with micro pattern gaseous detectors (MPGDs)~\cite{Titov2013,Guerard2012} is a promising option to achieve these requirements. Currently, the NMX detector consists of a natural gadolinium neutron converter mounted as cathode in a triple GEM detector~\cite{Altunbas2002} with a low material budget x/y strip readout. Since beam time at ESS will be a precious commodity, it is paramount to optimize the neutron detection efficiency and the signal to background ratio of the neutron detectors. To improve the neutron detection efficiency, an upgrade of the natural gadolinium converter to an enriched $^{157}$Gd converter is being considered. The relative improvement in detection efficiency of two samples of enriched $^{155}$Gd and $^{157}$Gd compared to natural gadolinium has been measured during two dedicated beam times at the EAR2 beamline of the n$\_$TOF experiment at CERN. In 2024, 39 hours were allocated to us from March 20th to March 22nd and 17 hours from October 24th to October 25th. 

\section{The NMX detector prototype} \label{sec:detector}
\begin{figure}[hbpt]
\centering
\subfloat[Gd-GEM 10~cm x 10~cm prototype\label{fig: Detector_Photo}]{
\includegraphics[width=.4\textwidth]{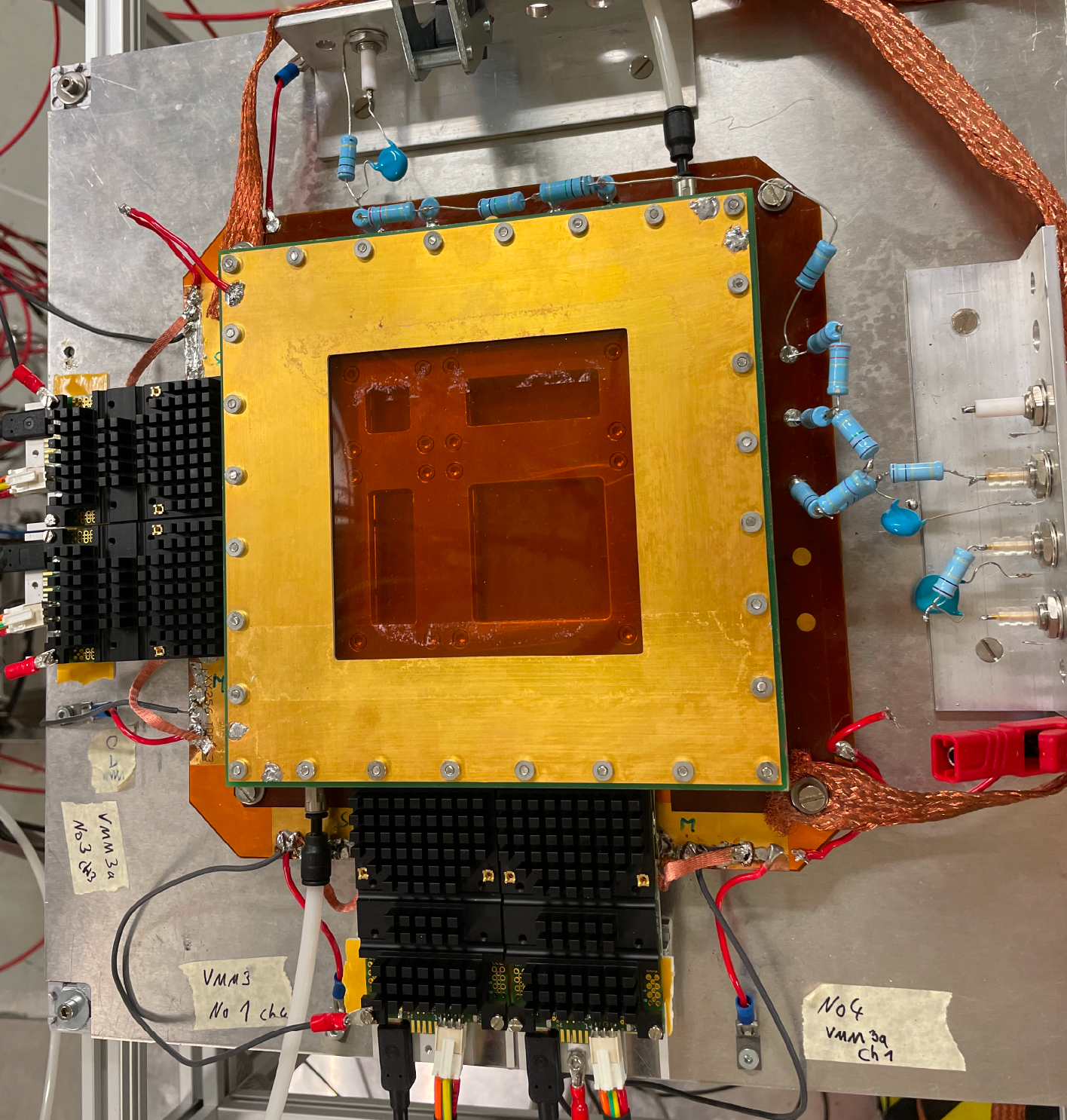}
}
\subfloat[Schematic view of Triple-GEM detector\label{fig: Detector_Scheme}]{
\includegraphics[width=.60\textwidth]{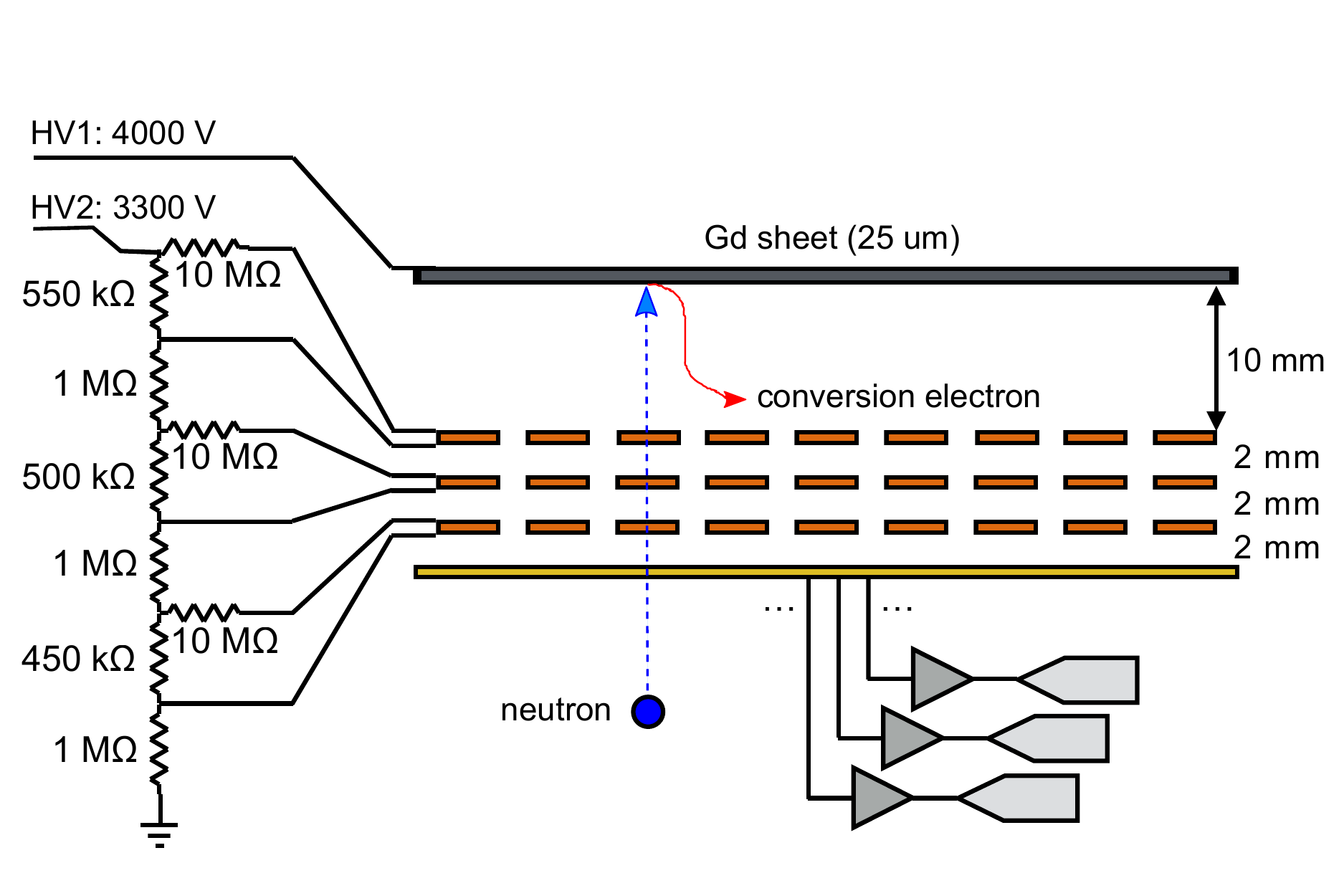}
}
\caption{Gd-GEM neutron detector}
\label{fig:Detector}
\end{figure}

For the measurements at n$\_$TOF, a smaller 10 x 10~cm$^{2}$ prototype was used in lieu of the full-size 51.2 x 51.2~cm$^{2}$ detector (Figure \ref{fig: Detector_Photo}). The conversion volume (or drift space) is 10~mm long, and a 25~$\mu$m gadolinium foil serves as cathode and neutron converter. A drift field of 700 V/cm was chosen to avoid electron attachment to electronegative impurities and the subsequent loss of primary ionization electrons, while keeping the drift velocity smaller than 2.0~cm/$\mu$s at atmospheric pressure. The three GEM foils~\cite{Sauli1997} in the detector are mounted at a distance of 2~mm from each other and powered with a resistor chain. The values of the resistors are stated in Figure \ref{fig: Detector_Scheme}. The detector operates at an effective gain of about 5000, and is usually flushed with Ar/CO$_2$ 70/30 mixture at 5~l/h at room temperature and atmospheric pressure. 

A schematic drawing of the NMX detector is displayed in Figure \ref{fig: Detector_Scheme}. 

The detector is operated in backwards mode. Neutrons traverse the low material budget detector readout and the three GEM foils, and then impinge on the natural Gd cathode. Natural Gd contains 14.8 $\%$ of $^{155}$Gd and 15.7 $\%$ of $^{157}$Gd. Both of these isotopes have a very high neutron capture cross section~\cite{Mastromarco2019}. Since enriched $^{157}$Gd and $^{155}$Gd foils are very expensive, it was not possible to obtain two cathode size converters. But fortunately two circular enriched Gd samples of 20 mm diameter with thicknesses of 40~$\mu$m ($^{155}$Gd) and 80~$\mu$m ($^{157}$Gd) could be borrowed from the n\_TOF collaboration at CERN. These foils were taped with copper tape to the inside of the 25~$\mu$m thick cathode of natural Gd in such a way that a good electrical connection was made (figure~\ref{fig: cathode_enrichedGd}). After the neutron capture, gadolinium releases prompt gamma particles with an energy of up to 9 MeV and conversion electrons with energies ranging from 29 keV to 250 keV. The signals resulting from these conversion electrons are used to determine the position of the neutron impact on the cathode with the help of the uTPC method~\cite{Pfeiffer2015, Pfeiffer2016}. The strip pitch of the x/y readout is 400 $\mu$m. There are 256 strips each in the \textit{x}- and \textit{y}-directions, which are read out with two RD51 VMM3a hybrids each in the \textit{x}-and \textit{y}-directions (figure \ref{fig: Detector_Photo}). Two VMM3a ASICs~\cite{DeGeronimo2012,Iakovidis2020} with 64 channels per chip are mounted on a RD51 VMM3a hybrid~\cite{Lupberger2018, Pfeiffer2022}, resulting in 128 channels per hybrid.

With this small prototype, a detection efficiency of 11.8$\%$ for neutrons with a wavelength of 2 \AA~\cite{Pfeiffer2016} was measured at IFE~\cite{IFE} in 2016 with a natural Gd converter. For that measurement, the detector used a traditional \textit{x/y} strip readout that contained an FR4 base~\cite{Bressan1999}.  We replaced this with a low material budget readout that uses a 100~$\mu$m thick Kapton foil to reduce the neutron scattering from $\sim 20~\%$ to $\sim 5~\%$. For a Gd-GEM detector with low material budget readout, as used in the n$\_$TOF measurements, the detection efficiency at 2.4~\AA~ was simulated with Geant4 10.1~\cite{Geant4a} to be around 15$\%$ with a threshold of 0 keV.

\section{Gadolinium neutron converters and self-shielding effects} \label{sec:converters}
\begin{figure}[hbpt]
\centering
\subfloat[Comparison of transmission and backward mode\label{fig: mode}]{
\includegraphics[width=.5\textwidth]{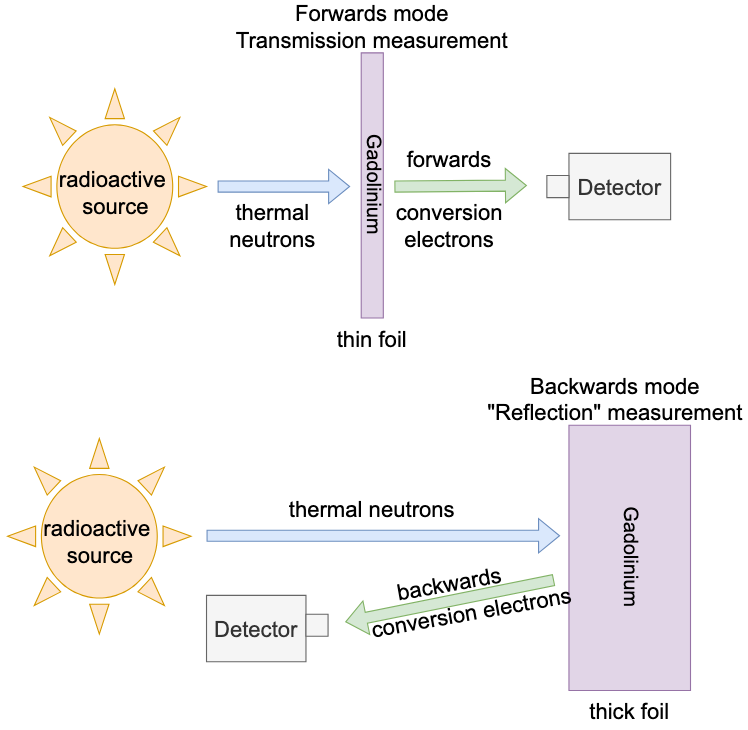}
}
\subfloat[Geant4 simulation of transmission and backwards mode\label{fig: Gd_probability}]{
\includegraphics[width=.5\textwidth]{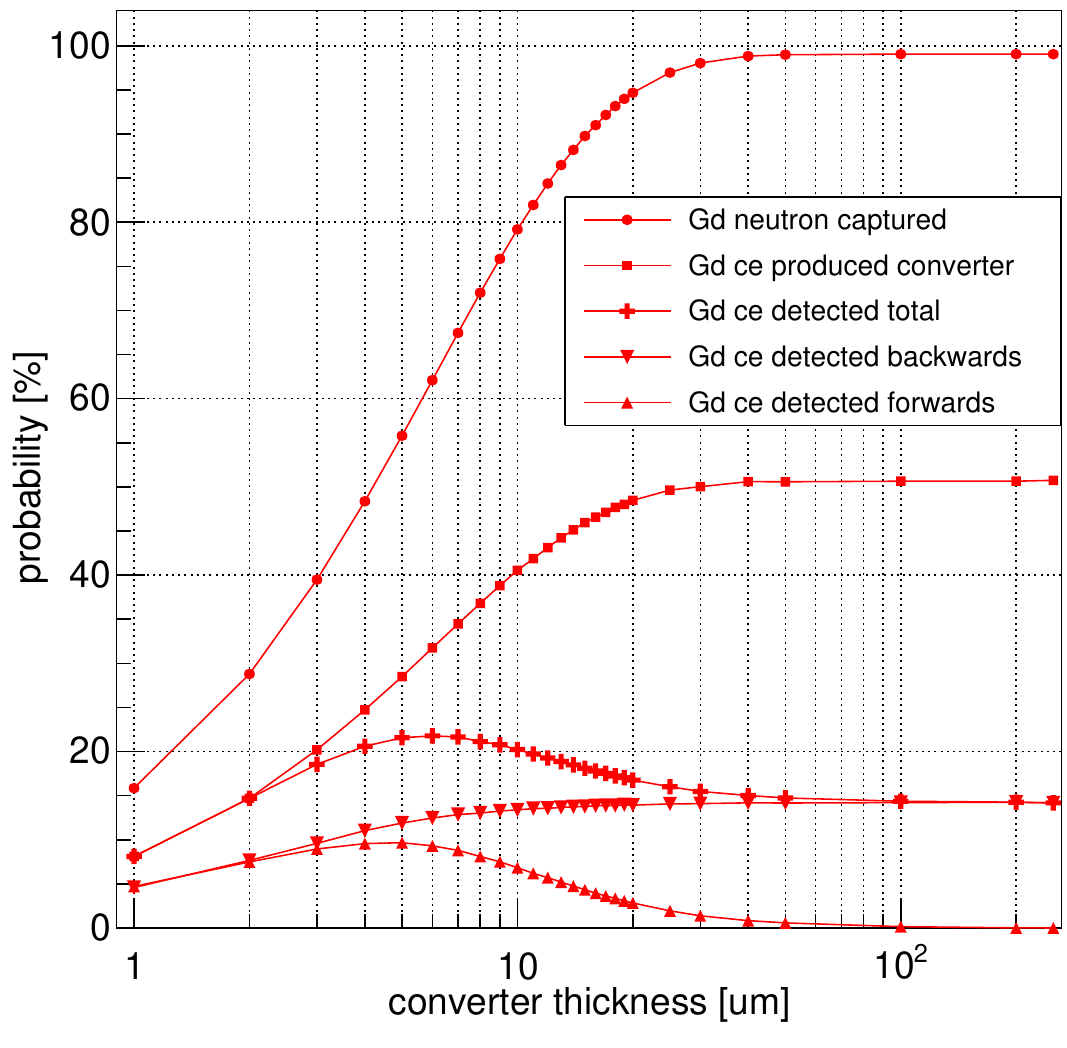}
}
\caption{Efficiency measurements of Gd converters can be carried out in forwards/transmission mode or backwards/reflection mode. In transmission mode the converter has to be very thin (around 5$\mu$m), whereas in backwards mode thicknesses of 25$\mu$m or larger are ideal for natural Gd. }
\label{fig:transmission}
\end{figure}

Since Gadolinium has a very high thermal neutron capture cross section (about 254000 barns for $^{157}$Gd and 60000 barns for $^{155}$Gd), in natural Gd 50$\%$ of the impinging thermal neutrons are already captured in the first 4$\mu$m of the converter. As the simulations with Geant4 10.1 for neutrons with a wavelength of 1.8 A in figure \ref{fig:transmission} show, at 10 $\mu$m thickness in natural Gd, 80$\%$ are captured, and the number increases to more than 99.5$\%$ for thicknesses 25$\mu$m and above. For enriched Gd the required thickness that stops all neutrons is even smaller. This means that for a stack of two Gd foils (enriched Gd on top of natural Gd), if the foil on the top is 40$\mu$m or thicker, the foil below basically does not see any neutrons. It is completely shielded by the foil on top. To optimize the detection efficiency of a detector, one has to maximise the number of conversion electrons that exit the converter and arrive in the gas volume of the detector. In transmission or forwards mode (figure~\ref{fig: mode}) the optimal thickness of a natural Gd converter is thin, a converter thickness of around 5$\mu$m leads to a maximum number of conversion electrons that traverse the Gd converter and reach the gas volume. For thicknesses larger than 5$\mu$m self-shielding effects occur in transmission mode (figure~\ref{fig: Gd_probability}). The conversion electrons (energies between 10 keV and 250 keV) are stopped in the Gd converter since their mean free path in Gd is short. 

In backwards (reflection) mode, which is used in the present work, the situation is different. When traversing the converter in the backwards direction (towards the incoming beam), conversion electrons produced in the first 5$\mu$m have the highest probability to escape into the gas volume. Increasing the converter thickness beyond about 10$\mu$m only marginally improves the efficiency, due to two types of self-shielding. First, neutron self-shielding: deeper regions of the converter see an increasingly attenuated neutron flux, because most neutrons are already captured in the first few micrometres. Second, electron self-shielding: conversion electrons created at larger depths have a rapidly decreasing probability of escaping, since they must traverse more material and are therefore stopped in the gadolinium. Nevertheless, in backwards (reflection) mode these two self-shielding mechanisms do not lead to a reduction of the number of extracted conversion electrons when the converter thickness is increased. Figure~\ref{fig: Gd_probability} illustrates that for natural Gd over 25$\mu$m only very small additional efficiency gains are obtained, the efficiency is essentially constant. There is no regime in which increased converter thickness leads to a decrease of the backwards electron yield. For enriched Gd the same applies, just are the required thicknesses smaller than for natural Gd.

For resonance neutrons with energies between 2 eV and 300 eV the cross sections are much lower with 1000 to 10000 barns at the resonant peaks and 10 to 100 barns for the baselines between the peaks. At resonance energies the 40$\mu$m or 80$\mu$m thick enriched Gd foils do not fully absorb all incident neutrons, therefore some neutrons reach the 25 $\mu$m  natural Gd support. However, even the highest-energy conversion electrons (about 250 keV) have an escape range much less than 40$\mu$m in Gd. They cannot traverse the 40$\mu$m or 80$\mu$m enriched foil, so the natural Gd support does not contribute conversion electrons to the measured signal. Capture gammas that undergo Compton scattering in the natural Gd underneath and around the enriched samples on the other hand can produce higher energy electrons that are detected in the detector.

\section{The experimental setup} \label{sec:setup}
\begin{figure}[hbpt]
\centering
\subfloat[NMX detector in setup at EAR2\label{fig: setup}]{
\includegraphics[width=.5\textwidth]{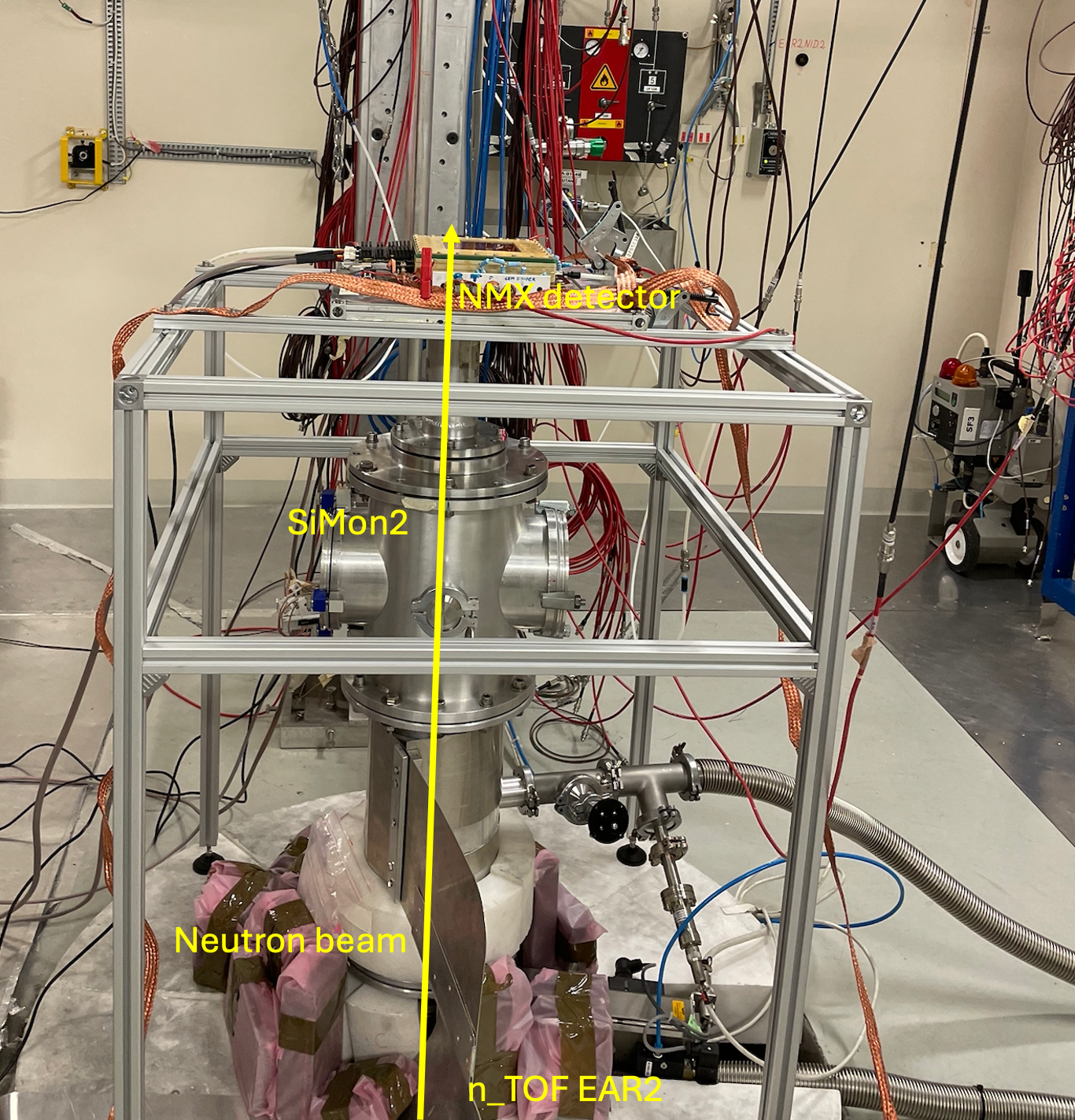}
}
\subfloat[Natural Gd cathode with enriched samples\label{fig: cathode_enrichedGd}]{
\includegraphics[width=.5\textwidth]{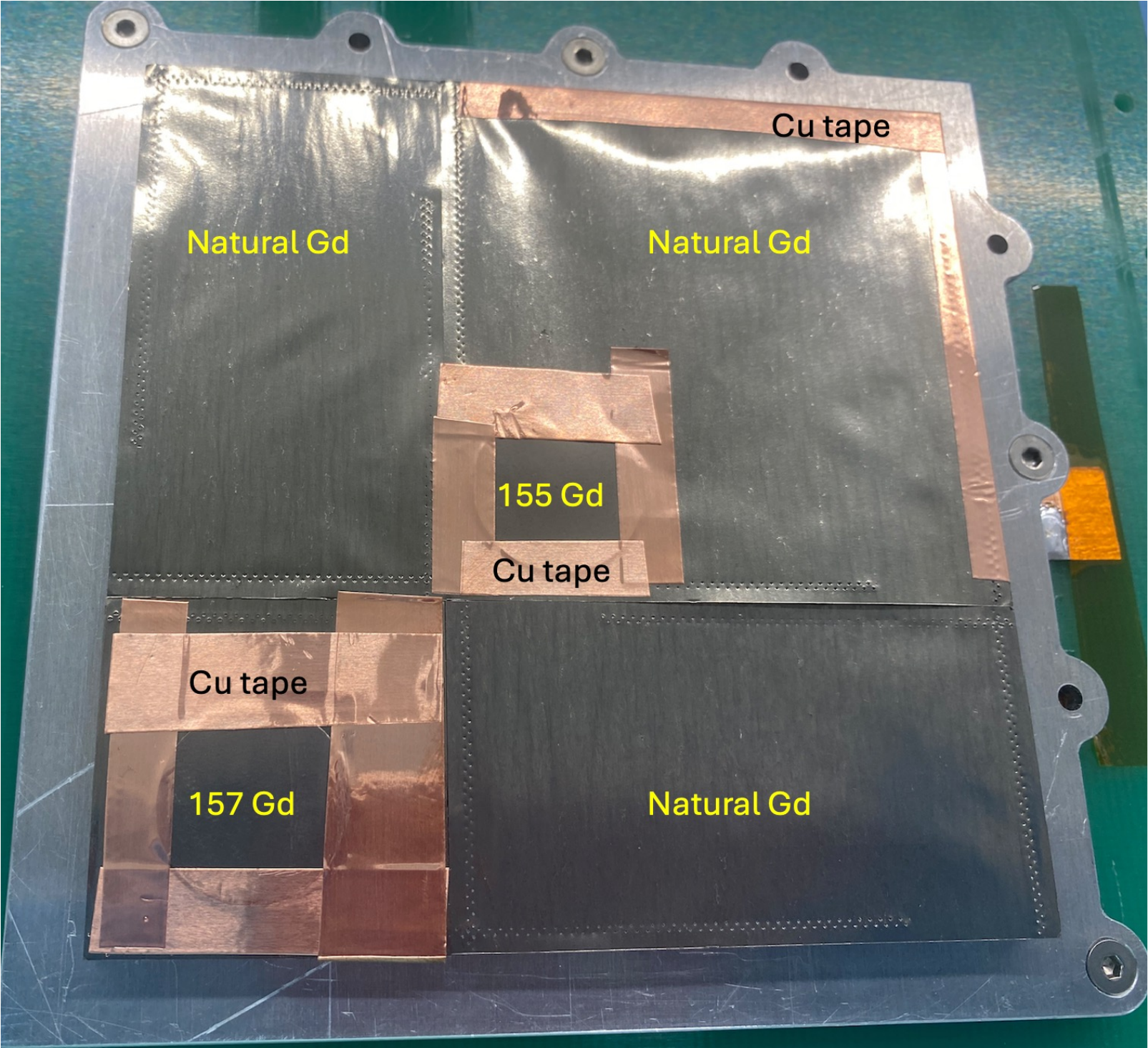}
}

\caption{Detector setup in the experimental cave of EAR2. The neutron beams moves in the vertical direction from the floor to the ceiling (left). Two samples of enriched $^{155}$Gd and $^{157}$Gd were installed in the NMX detector on the natural Gd cathode (right). }
\label{fig:Setup}
\end{figure}

The experimental setup at the n\_TOF EAR2 beam line is shown in Figure~\ref{fig: setup}. The n\_TOF experiment uses proton bunches that are first extracted from the PS accelerator, and subsequently hit a spallation target. Neutrons with energies between 10 meV and 1 GeV are delivered from the spallation target to the vertical EAR2 beam line, which has a neutron flight path of about 19 m. The NMX detector was installed at about 19.4 m from the center of the target on a support above the four n\_TOF SiMON2 beam monitors. These beam monitors are installed around the beam pipe and detect tritons produced from the interaction of neutrons with the thin $^{6}$LiF window inside the beam pipe. To measure separately the three different gadolinium regions, the detector was aligned in such a way that the beam was focused on one of the two samples or the natural Gd foil visible in Figure \ref{fig: cathode_enrichedGd}. After taking data for a few hours, the neutron beam was stopped and the detector moved. At the position of the detector (1.17 m above the floor) we measured a beam size of about 24 mm FWHM in x direction and 26 mm FWHM in y direction. At a height of 1.6 m above the floor a beam width of 21 mm FWHM was measured by the n\_TOF collaboration~\cite{Weiss2015}, we thus slightly overmeasure the width, probably by measuring part of the halo.

\begin{figure}[htbp]
\centering
\includegraphics[width=.90\textwidth]{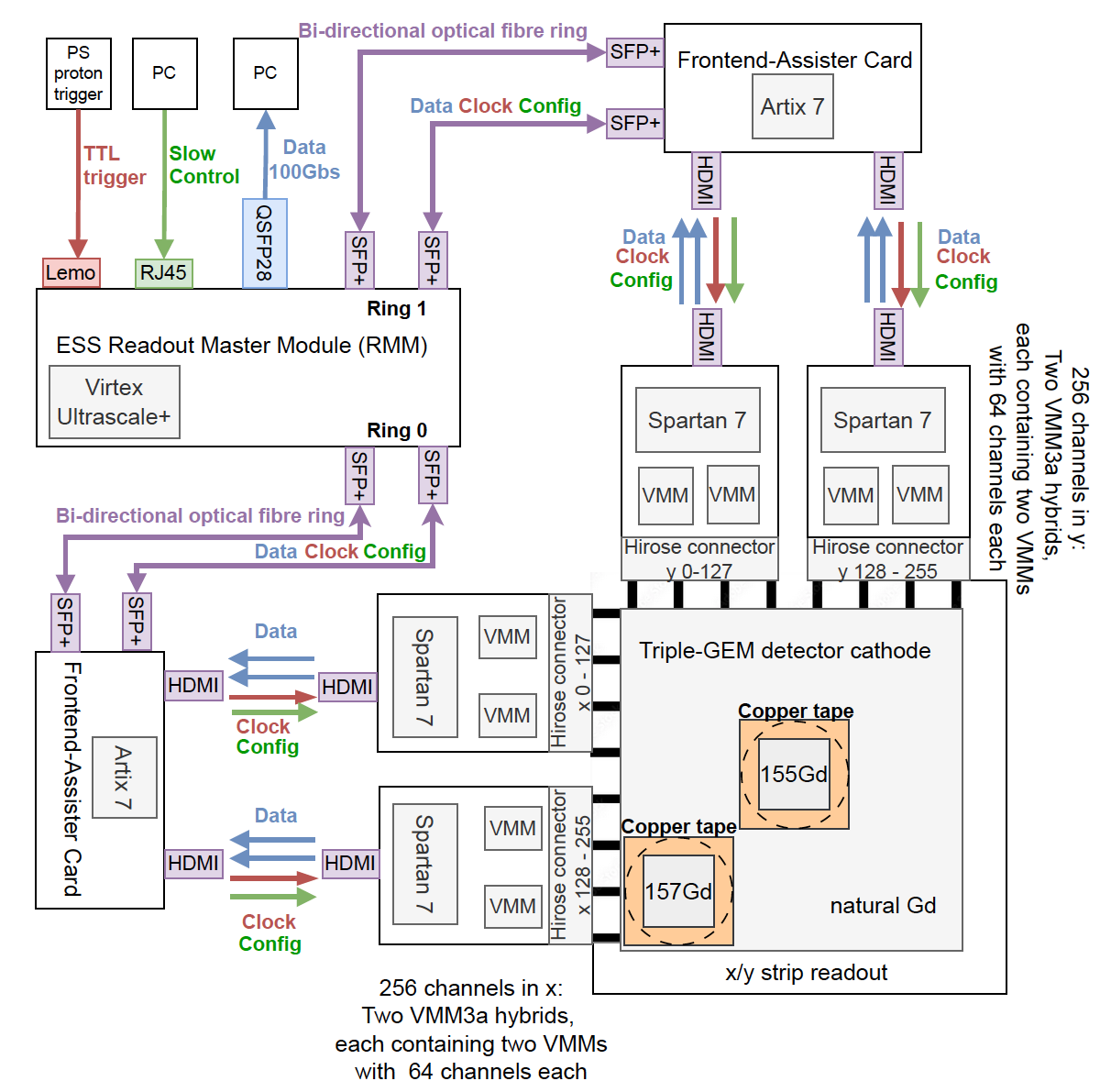}%
\caption{Schematic drawing of the detector and the readout system.}
\label{fig: readout}
\end{figure}

\section{The ESS readout system} \label{sec:readout}
The measurements at the n\_TOF EAR2 beam line at CERN described in this paper are the first time-of-flight experiment carried out with the NMX detector and the ESS readout system. A schematic drawing of the experimental setup is shown in figure~\ref{fig: readout}. The natural Gd cathode of the detector with the two circular enriched Gd samples of 20 mm diameter with thicknesses of 40~$\mu$m ($^{155}$Gd) and 80~$\mu$m ($^{157}$Gd) is displayed in figure~\ref{fig: cathode_enrichedGd}).

Two VMM3a hybrids were connected in x and y direction to the x/y strip readout of the detector. On each of the hybrids, the two VMM3a chips (64 channels per ASIC) digitize the electrical strip signals. The digitized signal is then transmitted to the Spartan 7 FPGA on the hybrid, which sends it 8b/10b encoded via LVDS over HDMI cable to the assister card. The assister cards serve as concentrators for the data from up to 6 hybrids. In the present experiment, two assister cards with two hybrids each were installed, one for the \textit{x}- and one for the \textit{y-} dimension. The assister cards convert the electrical signals into optical signals and send the data 8b/10b encoded over a bi-directional fibre ring with 6.6 Gbps to the ESS Readout Master Module (RMM). The RMM supports up to 12 bi-directional rings, which can have up to 16 assister nodes each. To optimize the bandwidth, two rings with only one node each were used. The RMM again is a concentrator module, and combines the data from all 12 rings into UDP jumbo frames transmitted via 100 Gbps to the data acquisition PC. On the PC, the complete data is written to disk. A fraction of the events is visualized to ensure the correct operation of the detector and readout system.

The DAQ PC also contains the slow control software that is used to configure the RMM, the assister cards, and the VMM3a hybrids. The slow control information is sent via an 1 Gbps ethernet connection to the RMM, which then passes the relevant slow control information on to each assister node on the 12 rings. The assisters again send the relevant slow control information to the hybrids. The clock is distributed the same way down from the RMM via the assisters to the hybrids. The RMM has the possibility to use an external time reference, but for the n\_TOF measurements the internal oscillator of the RMM was used. The PS proton trigger is fed into the TTL input of the RMM. The trigger signal appears then as trigger time stamp inside the readout data packages.

\section{Time-of-flight measurements} \label{sec:tof}
During the measurement campaign in March 2024, the intensity of the proton bunches that hit the spallation target was close to the nominal intensity of 7E+12 protons per bunch. At this intensity the neutron flux was too high for our detector, which was suffering from saturation effects. 

\begin{figure}[hbpt]
\centering
\subfloat[Time between low intensity bunches\label{fig: bunches_in_5minutes}]{
\includegraphics[width=.52\textwidth]{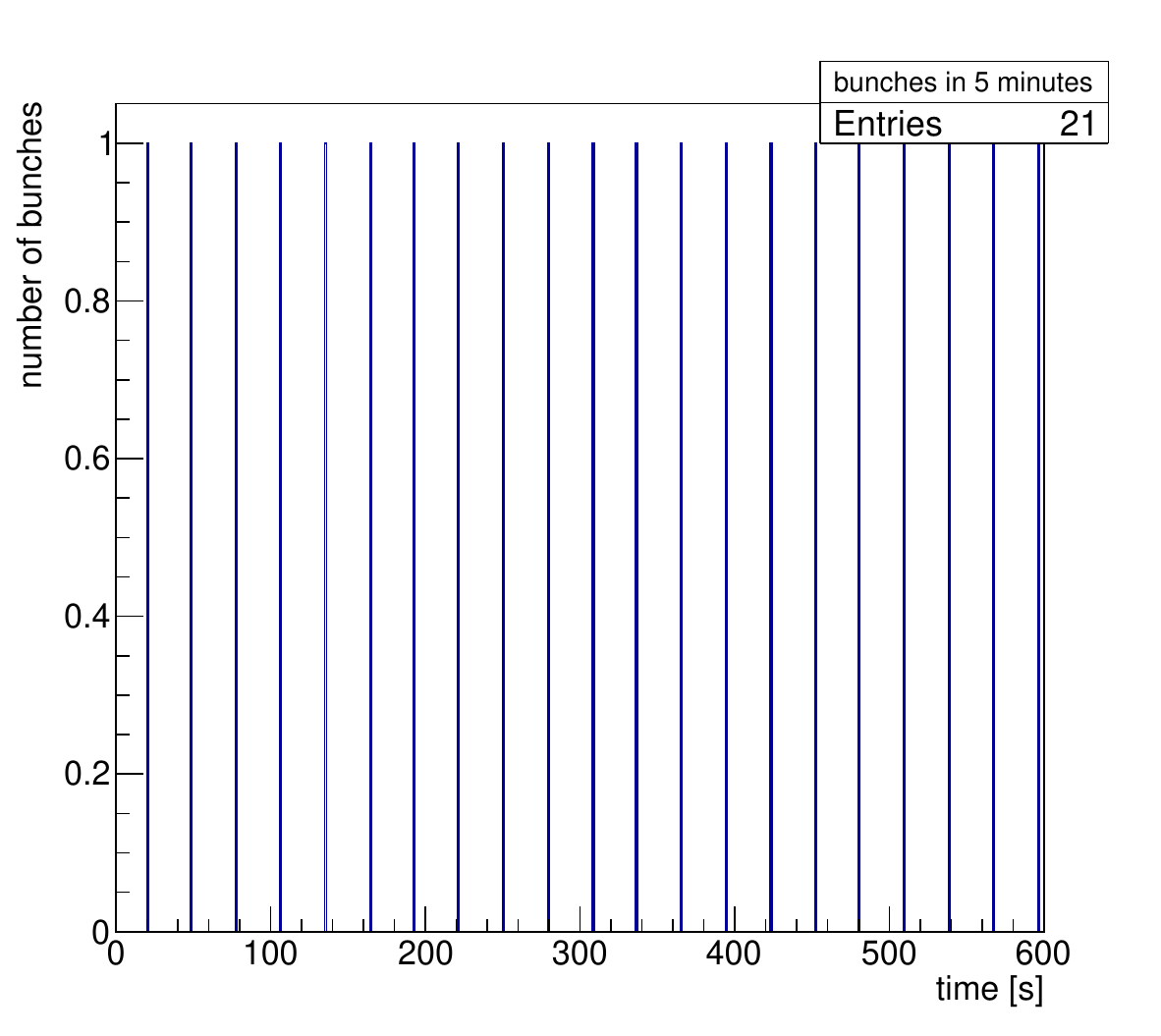}
}
\subfloat[Intensity of low bunches\label{fig: low_intensity_bunches}]{
\includegraphics[width=.48\textwidth]{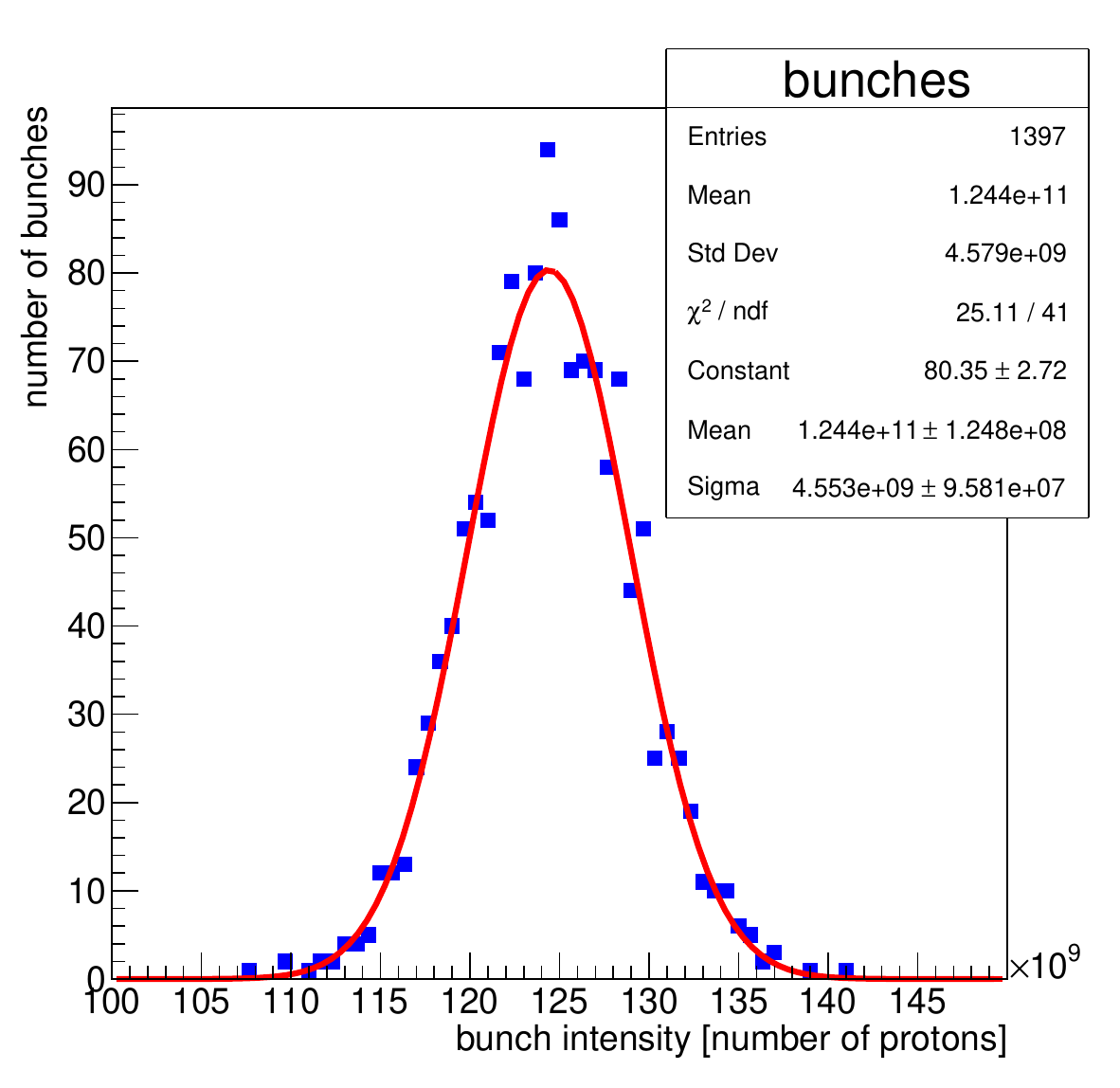}
}
\caption{October 2024: Frequency and intensity of low intensity bunches (around 1.25E+11 protons). If there are low intensity bunches between the high intensity bunches, they appear about every 30 s. Higher intensity bunches, and especially the full intensity bunches of more than 7.0E+12 protons, produce too many neutrons for our detector and lead to saturation. Bunches with intensities larger than 2E+11 protons have thus been excluded from our analysis. On average there was a protons bunch every 4s during our experiment in October. }
\label{fig:Bunches}
\end{figure}

In October 2024, however, in addition to the nominal intensity bunches, smaller bunches with an intensity of around 1.25E+11 protons were regularly injected into the PS. Whereas on average every 4 seconds a bunch of all intensities was hitting the target, the frequency of the low intensity bunches was much lower, with one bunch every 30 seconds, as shown in Figure~\ref{fig: bunches_in_5minutes}. To normalize our neutron intensity measurements, the proton intensity responsible for the neutron production has to be known. The time and the intensity of the extracted proton bunches was provided to us in the form of a root tree by the n\_TOF experiment. We then matched the Unix time stamp in nanoseconds of the proton bunch to our proton trigger time stamp with 11.36 ns resolution. Since we were relying on the internal oscillator of the RMM as time reference, during the successful matching a time drift of 6 ns per second (6 ppm) was observed between our recorded trigger time stamp and the time stamp in the root tree.

\begin{figure}[hbpt]
\centering
\includegraphics[width=.90\textwidth]{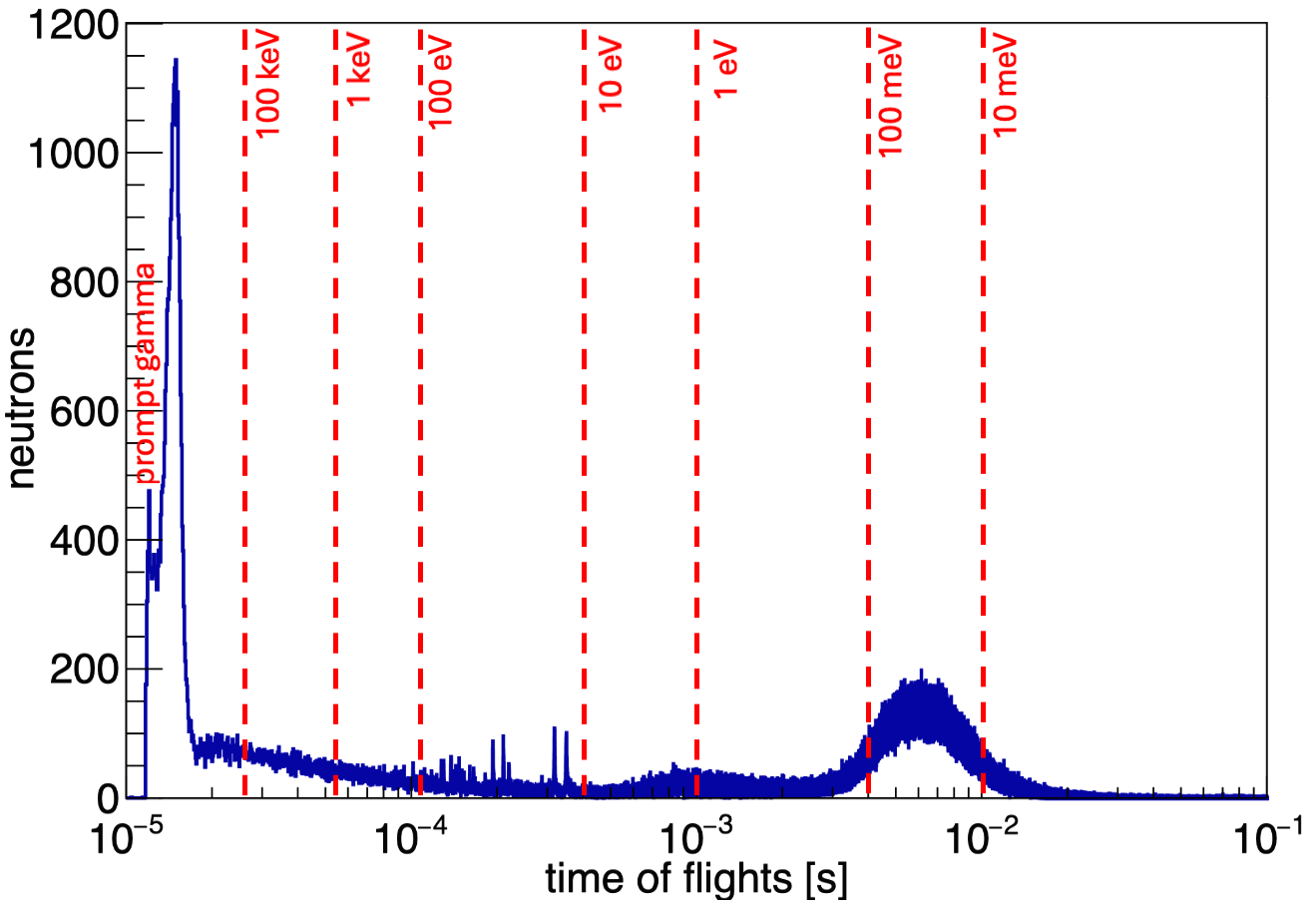}%
\caption{Time-of-flight measurements of neutrons detected on the enriched $^{157}$Gd sample. On the left of the plot one can see the prompt gamma peak, which occurs at about 12 us. The prompt gamma time-of-flight represents the t0 for the calculation of the neutron energy. }
\label{fig: tof_clusters}
\end{figure}

In Figure~\ref{fig: tof_clusters} the time-of-flight of neutron events detected on the enriched $^{157}$Gd sample is shown. The prompt gammas appear at about 12 $\mu$s. This t$_{0}$ time serves subsequently as time correction to convert the time-of-flight into energy. To convert the time-of-flight to energy, the following relativistic equations can be used:

\begin{align}
\mathbf{measured~time~of~flight}~tof \\
\mathbf{length~flight~path}~l_{path} &= 19.4~m \\
\mathbf{time~correction}~t_{cor} &= 12~\mu s \\
\mathbf{speed~of~light}~c &= 299792458~\frac{m}{s} \\
\mathbf{neutron~mass}~m &= 939.56542052 \cdot 10^{6}~\frac{eV}{c^2} \\
\gamma &= \frac{1}{\sqrt{1 - (\frac{l_{path}}{(tof-t_{cor}) \cdot c})^{2}}} \\
\mathbf{relativistic~kinetic~energy}~E_{kin} &= (\gamma-1)\cdot m \cdot c^{2}~eV \\
\end{align}
   
For neutrons with energies below a few keV, the non-relativistic equations suffice:    
\begin{align}
\mathbf{kinetic~energy}~E_{kin} &= \frac{1}{2} \frac{m}{c^2} v^{2}~eV \\
&= \frac{1}{2}  \frac{m}{c^2} (\frac{l_{path}}{tof-t_{cor}})^{2}~eV \\
\end{align} 

\begin{figure}[hbpt]
\centering
\includegraphics[width=0.9\textwidth]{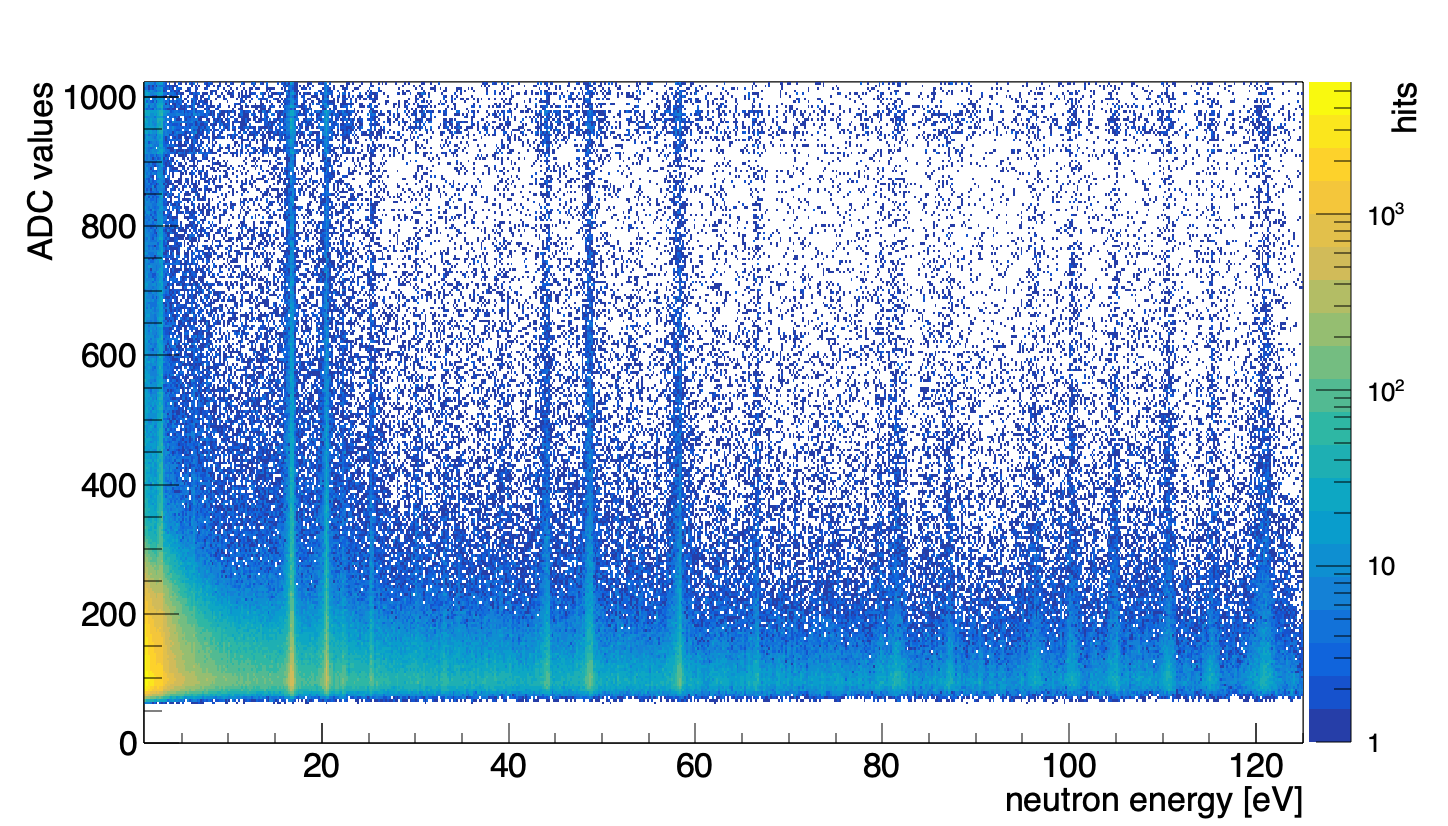}%
\caption{Resonant peaks between 1 eV and 125 eV visible in the ADC distribution measured with $^{157}$Gd. }
\label{fig: Resonance_ADC}
\end{figure}

Two regions of the time-of-flight spectrum will be analyzed in more detail, the resonant region and the thermal neutron peak. The region with neutron energies between 1 eV and 125 eV shows numerous resonant peaks that can be clearly distinguished in the ADC distribution of the VMM3a channels displayed in figure~\ref{fig: Resonance_ADC}.

\newpage

\begin{figure}[hbpt]
\centering
\includegraphics[width=1.0\textwidth]{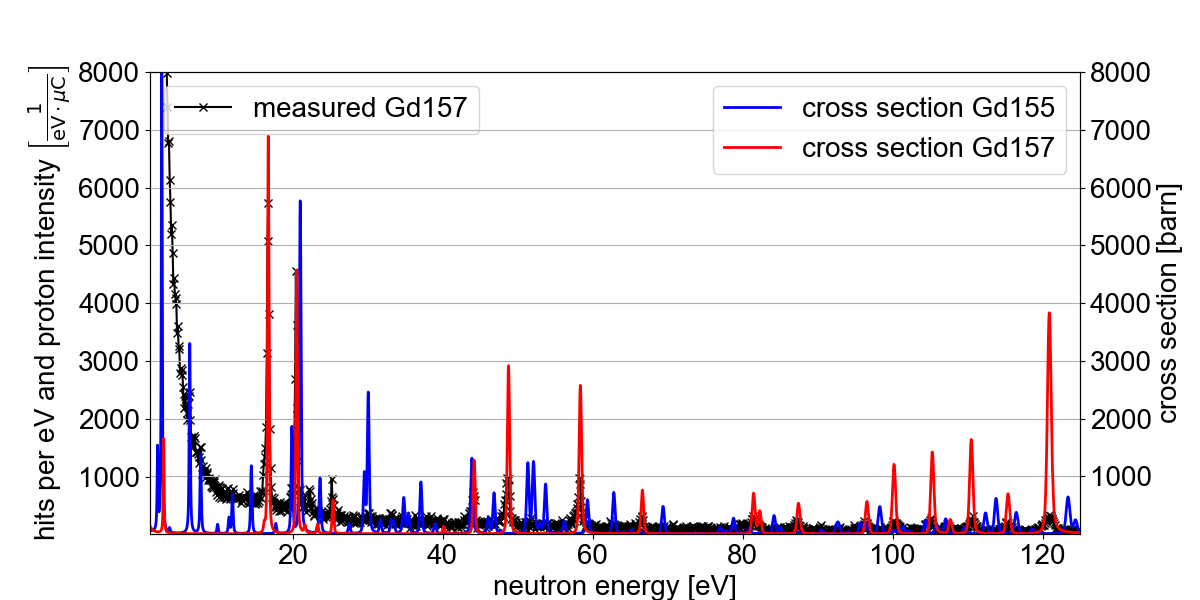}%
\caption{Hit distribution per eV normalized with the proton current as measured on the $^{157}$Gd sample. On the second y-axis at the right side the ENDF VIII.0 total neutron cross sections for $^{157}$Gd and $^{155}$Gd can be seen. }
\label{fig: Resonance_Gd157}
\end{figure}

When using 19.4 m as flight path and 12 $\mu$s as time correction, the measured peaks do not perfectly overlap with the cross section peaks of $^{157}$Gd. To find the exact time correction and flight path that makes the peaks overlap, two peaks in the ENDF VIII.0 $^{157}$Gd cross section plot (figure~\ref{fig: Resonance_Gd157}) were chosen.  Their respective time-of-flight (tof$_{1}$ for the first peak and tof$_{2}$ for the second peak) and energy (E$_{1}$ and E$_{2}$) were subsequently put into the following system of equations, which is then solved for t$_{cor}$ and l$_{path}$.

\begin{align}
E_{kin} &= \frac{1}{2} \frac{m}{c^2} v^{2}~eV \\
&= \frac{1}{2}  \frac{m}{c^2} (\frac{l_{path}}{tof-t_{cor}})^{2}~eV \\
l_{path} &= (tof_{1}-t_{cor}) \cdot \sqrt{\frac{2E_{1}}{m}} \cdot c \\
 &= (tof_{2}-t_{cor}) \cdot \sqrt{\frac{2E_{2}}{m}} \cdot c \\
K_{1} &= \sqrt{\frac{2E_{2}}{m}} \cdot c \\
K_{2} &= \sqrt{\frac{2E_{1}}{m}} \cdot c \\
t_{cor} &= \frac{tof_{2} \cdot K_{2} - tof_{1} \cdot K_{1}}{K_{2} - K_{1}}\\
\end{align}

The results for four different pairs of peaks are shown in table~\ref{table:path} below.

\begin{table}[h!]
\centering
\begin{tabular}{ |c|c|c|c|c|c| } 
\hline
Peak1 Energy & Peak 1 ToF & Peak2 Energy & Peak 2 ToF & t$_{cor}$ & l$_{path}$ \\
\hline
16.77 eV & 350.3 $\mu$s & 120.90 eV & 137.8 $\mu$s & 11.7 $\mu$s & 19.18 m \\
\hline
16.77 eV & 350.3 $\mu$s & 137.90 eV & 129.9 $\mu$s & 11.8 $\mu$s & 19.17 m \\
\hline
16.77 eV & 350.3 $\mu$s & 143.55 eV & 127.5 $\mu$s & 11.7 $\mu$s & 19.18 m \\
\hline
16.77 eV & 350.3 $\mu$s & 239.32 eV & 101.5 $\mu$s & 11.9 $\mu$s & 19.17 m \\
\hline
\end{tabular}
\caption{Calculated t$_{cor}$ and l$_{path}$ from energy and time-of-flight of cross-section peaks.}
\label{table:path}
\end{table}

Whereas the calculated time correction t$_{cor}$ agrees with the measured value of the prompt gamma flash, the calculated flight path of 19.18~m is already about 1.1$\%$ shorter than the assumed geometrical distance of 19.4~m. The resonance energies measured by the n\_TOF collaboration~\cite{Mastromarco2019} differ with e.g. E$_{1}$=16.7946~eV and E$_{2}$=120.861 eV from the ENDF VIII.0 values. Using the n\_TOF collaboration values, one arrives at t$_{cor}$= 11.35~$\mu$s, and l$_{path}$=19.21~m. More precisely, the calculated flight path l$_{path}$  consists of the geometrical distance L$_0$ and an energy dependent part, the moderation length $\lambda$. For the purpose of this paper, and when looking at energy ranges between 1eV and 200 eV, it suffices to use a $\lambda$ that is constant in this range. Taking a moderation length in the order of a few centimeters, the actual geometrical distance is even shorter than 19.18 m.~\footnote{For a spallation target like the one at n\_TOF, the moderation length is a statistical measure of the additional distance neutrons effectively travel within the target–moderator system before escaping with thermal or epithermal energies. Its precise value depends on neutron energy and geometry and can only be reliably determined through Monte Carlo simulation.}. Table~\ref{table:conversion} contains the conversion of energy and time-of-flight to wavelength as reference.

Figure~\ref{fig: Resonance_Gd157} uses the re-calculated l$_{path}$ and t$_{cor}$. It shows in the energy range between 1eV and 125 eV the ENDF VIII.0 total neutron cross sections~\cite{ENDF} for $^{157}$Gd and $^{155}$Gd. The measured data were obtained by converting the time-of-flight of each VMM3a channel into an energy for the run when the neutron beam was focused on the $^{157}$Gd. As expected, the measured peaks agree with the large $^{157}$Gd cross section peaks. The $^{157}$Gd measurement lasted only 10 hours, and due to detector saturation only 833 bunches with an average intensity of 1.24E+11 protons could be included in the analysis. Therefore not enough statistics were acquired to find agreement with the smaller $^{157}$Gd cross section peaks. On the other hand, for very high $^{155}$Gd cross section peaks, one also sees small peaks in the measured $^{157}$Gd data. There are two reasons for this. First, the analyzed region does not exclusively consist of $^{157}$Gd, but also includes on the outside away from the center of the beam a few mm$^{2}$ of natural Gd. Second, the sample of enriched $^{157}$Gd contains a 0.29 $\%$ contamination of $^{155}$Gd\cite{Mastromarco2019}.

\begin{table}[h!]
\centering
\begin{tabular}{ |c|c|c| } 
\hline
Energy [eV]  & ToF [ms] & Wavelength [A]\\
\hline
0.01 & 13.88 & 2.860 \\
\hline
0.025 & 8.78 & 1.809 \\
\hline
0.05 & 6.21 & 1.279 \\
\hline
0.075 & 5.08& 1.044 \\
\hline
0.1 & 4.40 & 0.905 \\
\hline
0.5 & 1.97 & 0.405 \\
\hline
1.0 & 1.40 & 0.286 \\
\hline
10.0 & 0.45 & 0.091 \\
\hline
20.0 & 0.32 & 0.064 \\
\hline
30.0 & 0.27 & 0.052 \\
\hline
40.0 & 0.23 & 0.045 \\
\hline
50.0 & 0.21 & 0.041 \\
\hline
60.0 & 0.19 & 0.037 \\
\hline
70.0 & 0.18 & 0.034 \\
\hline
80.0 & 0.17 & 0.032 \\
\hline
90.0 & 0.16 & 0.030 \\
\hline
100.0 & 0.15 & 0.029 \\
\hline
110.0 & 0.14 & 0.027 \\
\hline
120.0 & 0.14 & 0.026 \\
\hline
\end{tabular}
\caption{Conversion of energy and time-of-flight and neutron wavelength.}
\label{table:conversion}
\end{table}

\begin{figure}[htbp]
\centering
\subfloat[$^{157}$Gd\label{fig: Gd157_resonances}]{
\includegraphics[width=.8\textwidth]{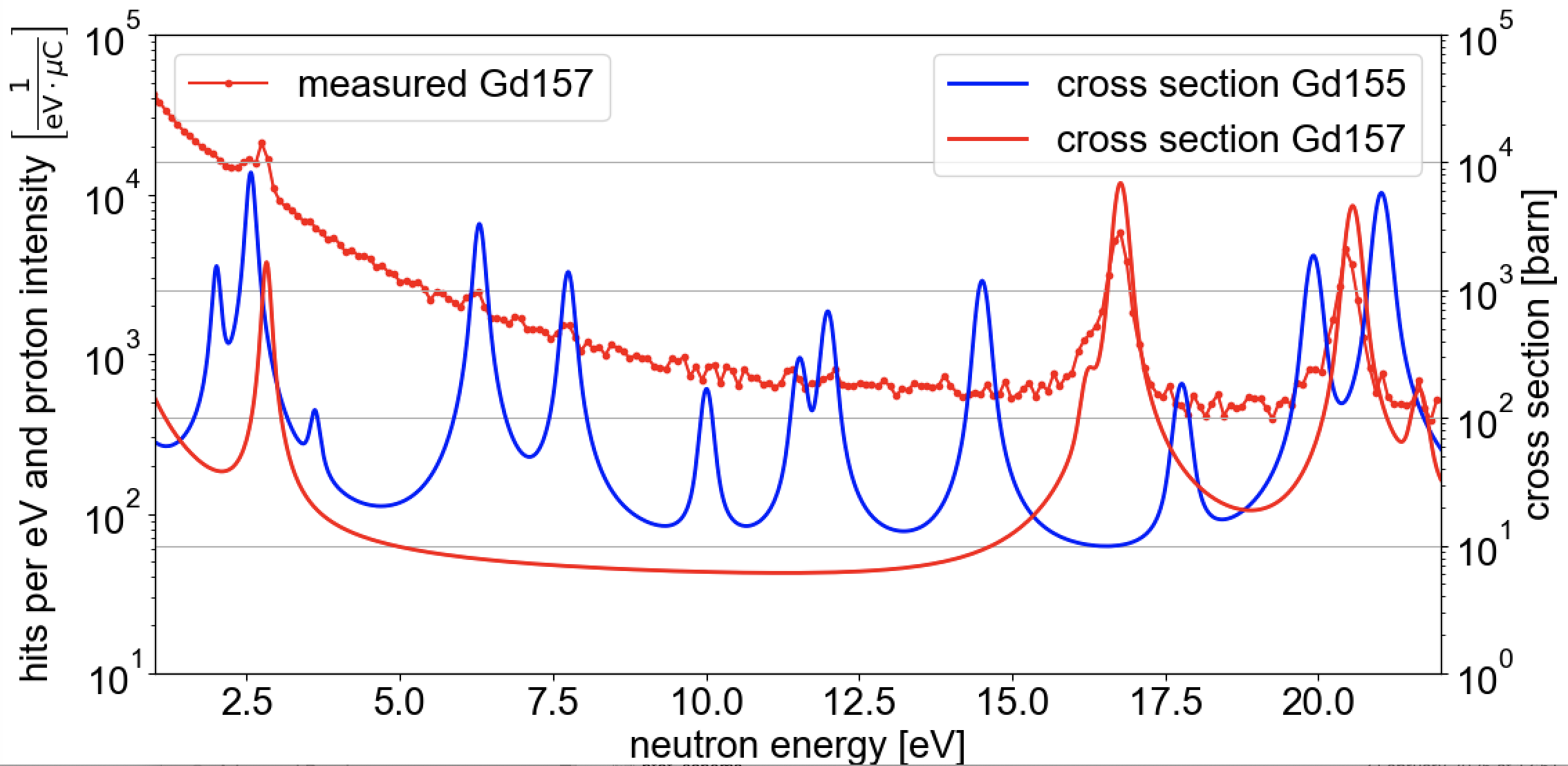}%
}
\\
\subfloat[$^{155}$Gd\label{fig: Gd155_resonances}]{
\includegraphics[width=.8\textwidth]{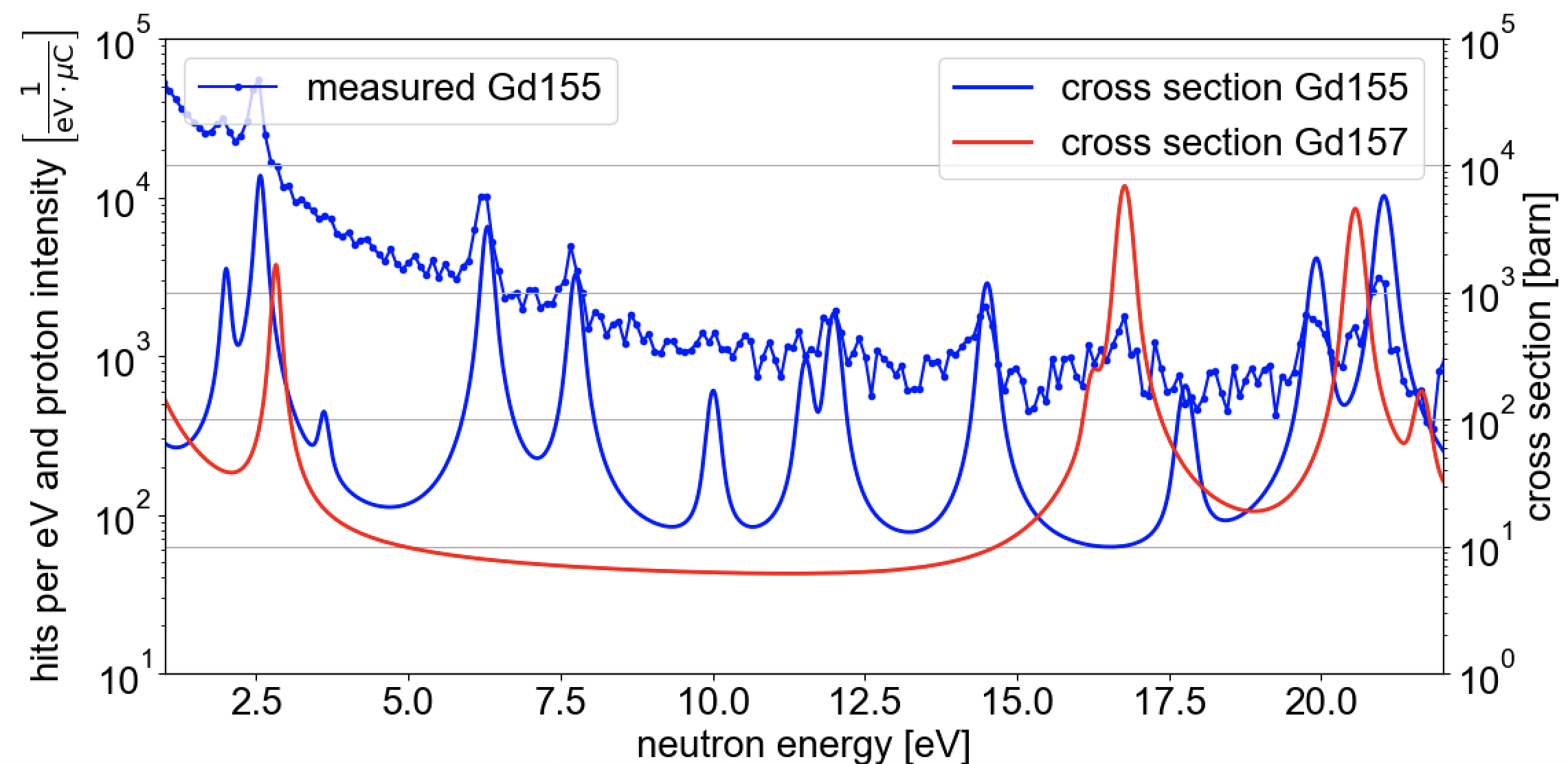}%
}
\\
\subfloat[natural Gd\label{fig: natGd_resonances}]{
\includegraphics[width=.8\textwidth]{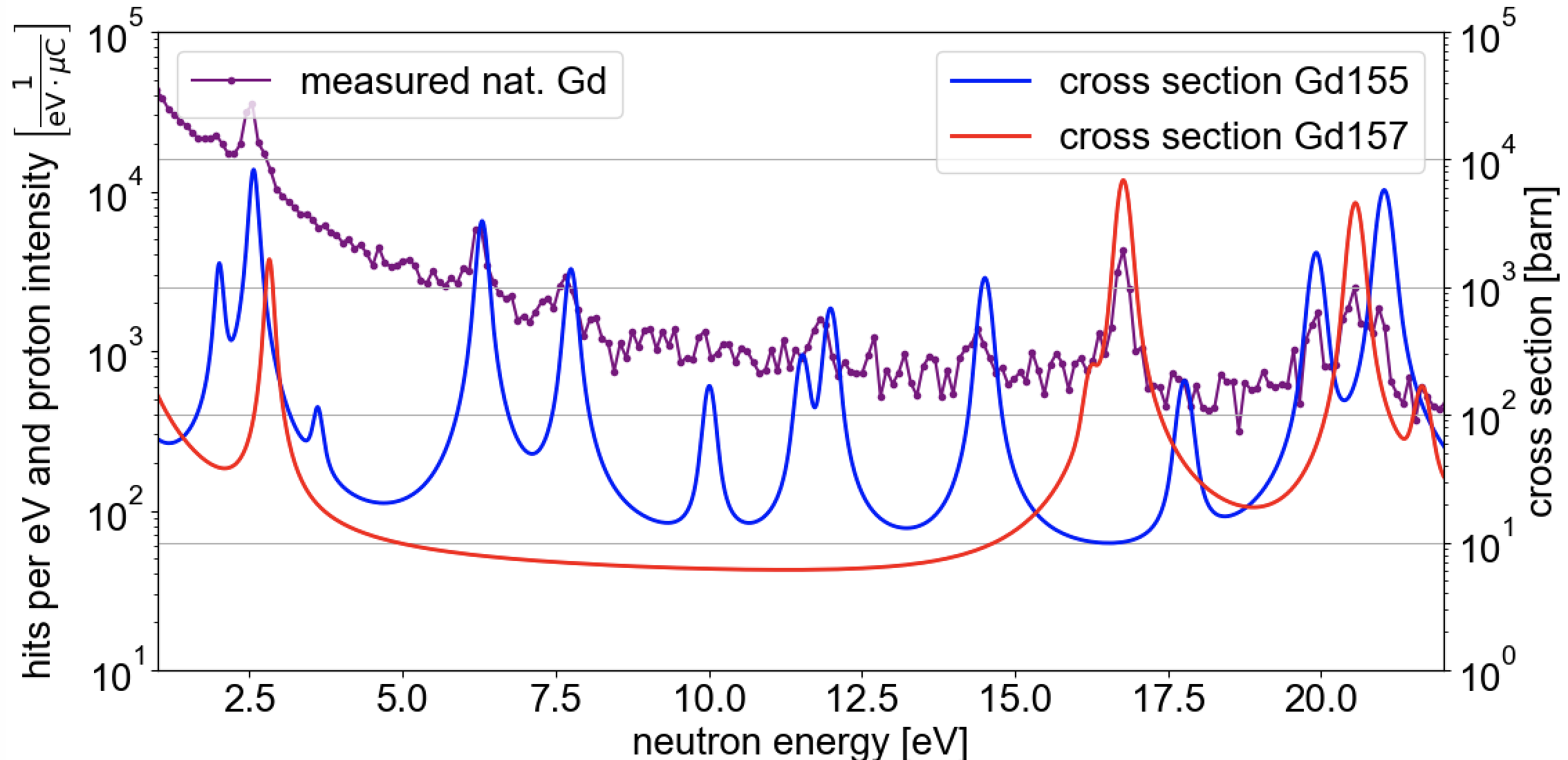}%
}
\caption{Comparison of the ENDF VIII.0 cross sections with the measured data in the three regions: $^{157}$Gd (a), $^{155}$Gd (b) and natural Gd (c)}
\label{fig: comparison_resonances}
\end{figure}

Figure~\ref{fig: comparison_resonances} compares the ENDF VIII.0 cross sections with the data measured on the different Gd regions. As expected the measured resonances for $^{157}$Gd follow the $^{157}$Gd ENDF VIII.0 resonances, and the measured $^{155}$Gd resonances agree with the $^{155}$Gd ENDF VIII.0 peaks. The strong resonance in the $^{157}$Gd ENDF VIII.0 at 16.77~eV is also weakly seen in the measured $^{155}$Gd data. As in the case of the $^{157}$Gd sample, the analyzed region for $^{155}$Gd also contains at the outside, away from the center of the beam, several mm$^2$ of natural Gd. Further, the $^{155}$Gd sample is contaminated with 1.14 $\%$ of $^{157}$Gd. The measured resonances in the natural Gd region follow both the $^{155}$Gd and the $^{157}$Gd ENDF VIII.0 resonances, but with reduced amplitudes compared to the enriched samples.

\section{Comparison of the efficiencies of $^{157}$Gd, $^{155}$Gd and natural Gd} \label{sec:efficiency}
\begin{figure}[htbp]
\centering
\includegraphics[width=1.0\textwidth]{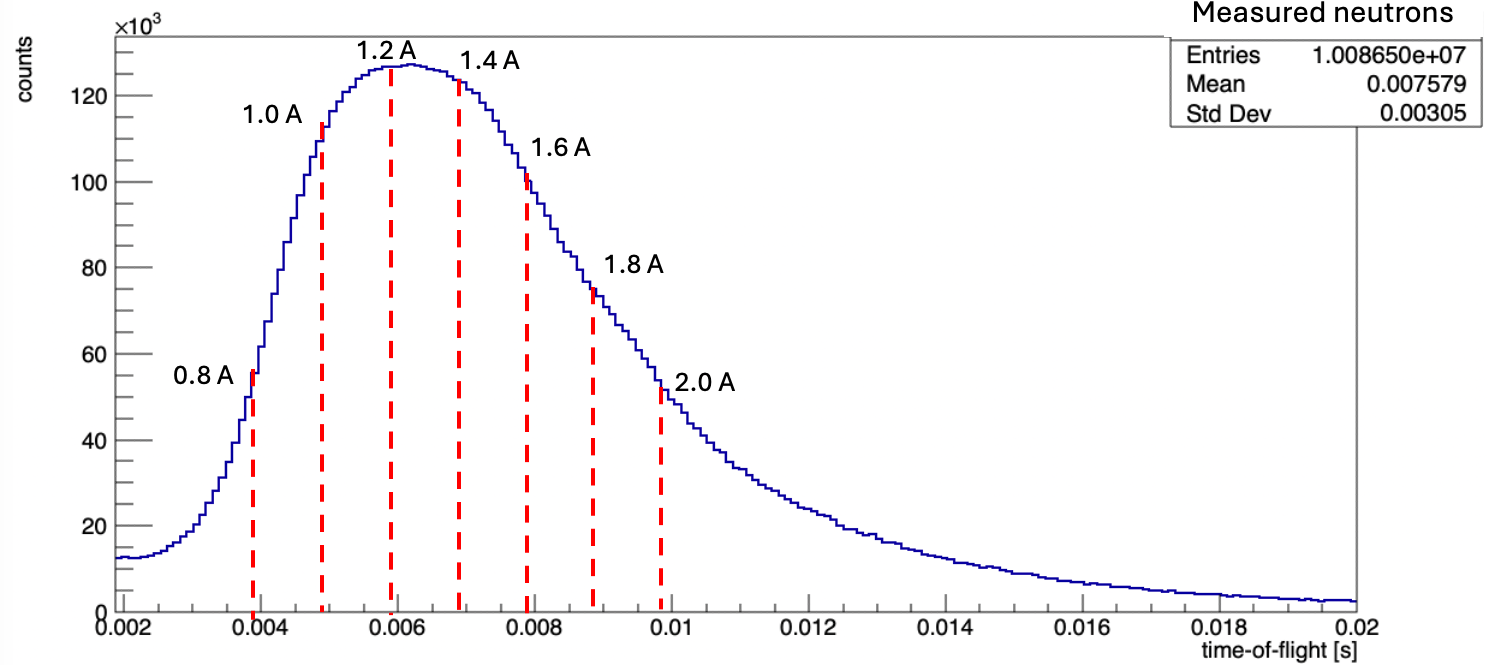}%
\caption{Thermal neutron peak as measured with the $^{157}$Gd sample. The largest intensity was measured in the wavelength range between 0.8 A and 2.0 A.  }
\label{fig: Thermal Peak}
\end{figure}

\begin{figure}[htbp]
\centering
\subfloat[natural Gd - 0.8A\label{fig: natGd_0p8A}]{
\includegraphics[width=.33\textwidth]{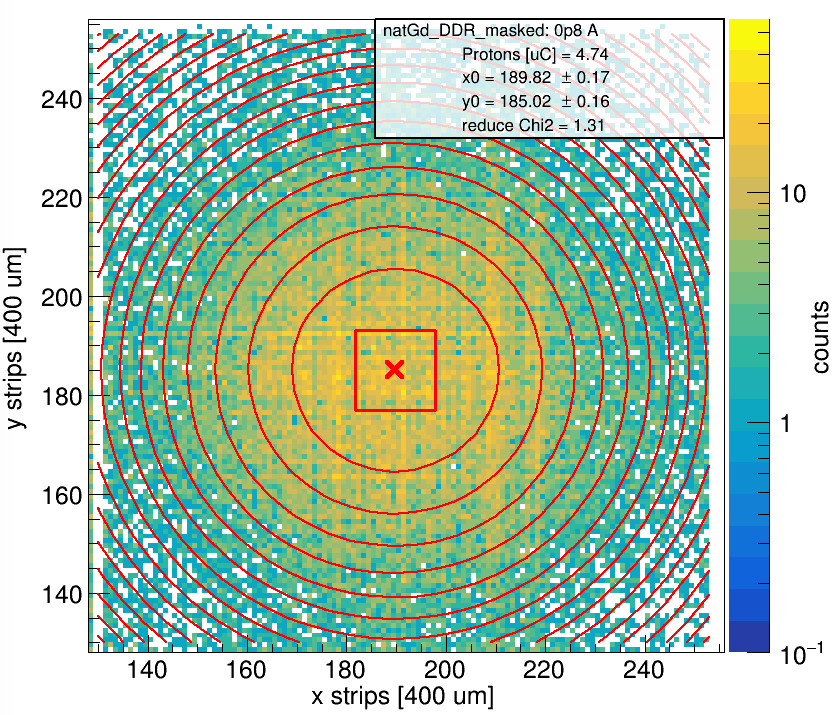}%
}
\subfloat[$^{157}$Gd - 0.8A\label{fig: Gd157_0p8A}]{
\includegraphics[width=.33\textwidth]{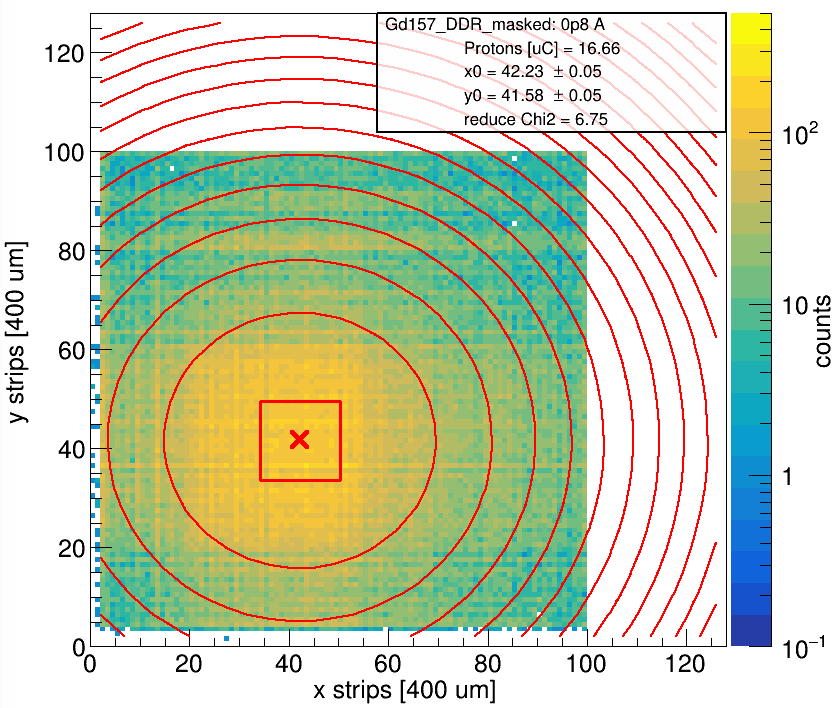}%
}
\subfloat[$^{155}$Gd - 0.8A\label{fig: Gd155_0p8A}]{
\includegraphics[width=.33\textwidth]{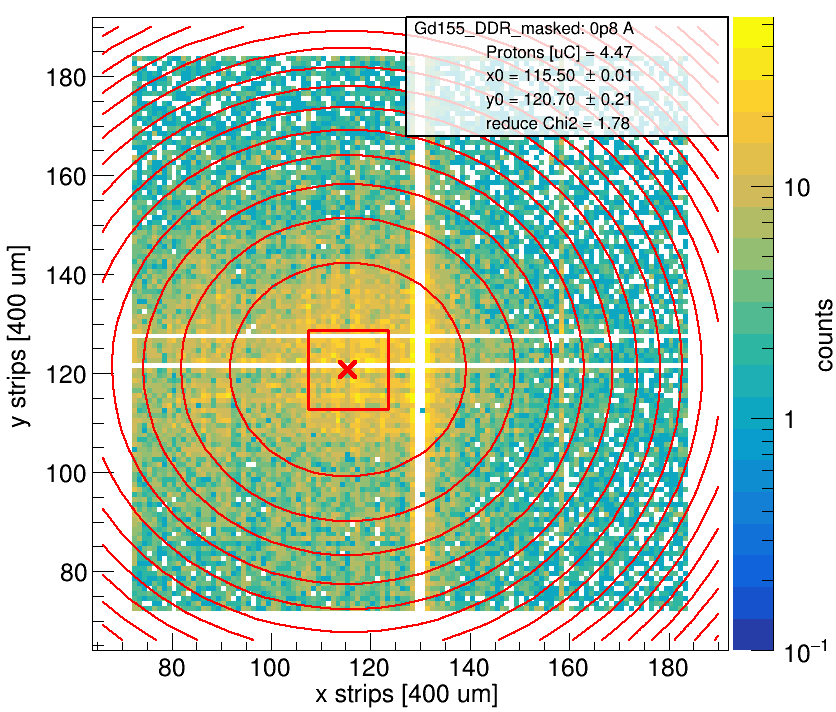}%
}
//
\subfloat[natural Gd - 1.4A \label{fig: natGd_1p4A}]{
\includegraphics[width=.33\textwidth]{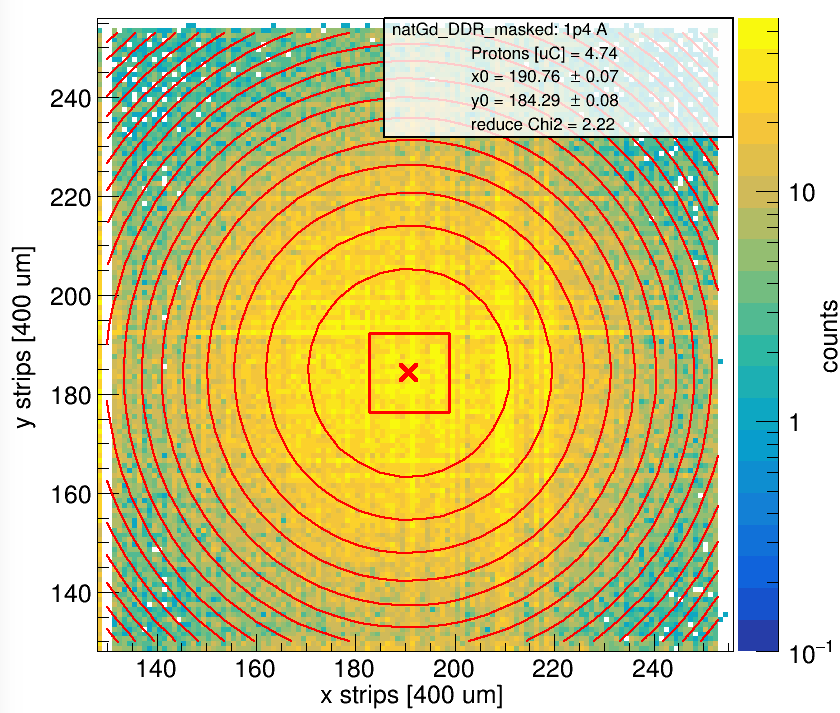}%
}
\subfloat[$^{157}$Gd - 1.4A\label{fig: Gd157_1p4A}]{
\includegraphics[width=.33\textwidth]{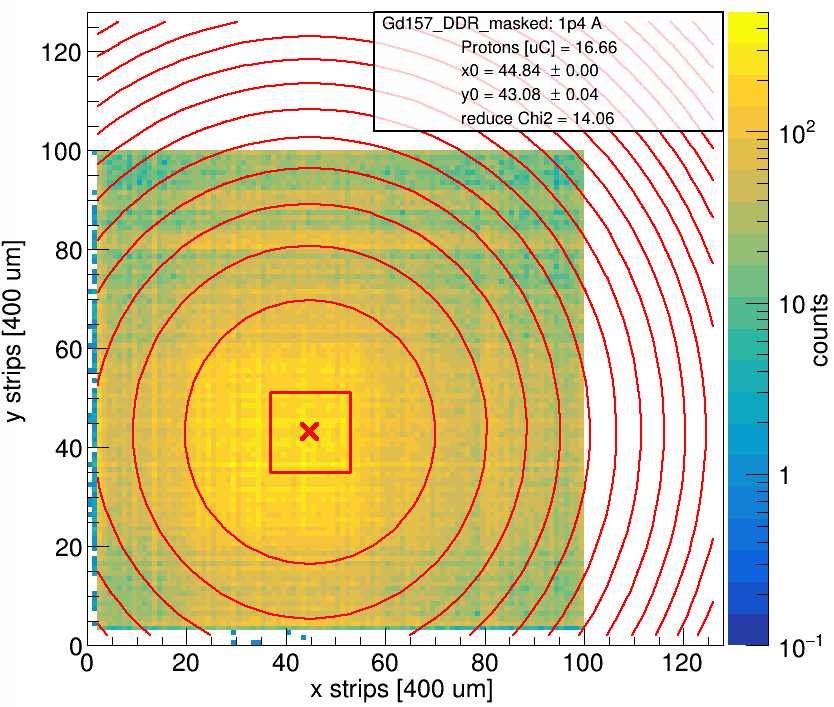}%
}
\subfloat[$^{155}$Gd - 1.4A\label{fig: Gd155_1p4A}]{
\includegraphics[width=.33\textwidth]{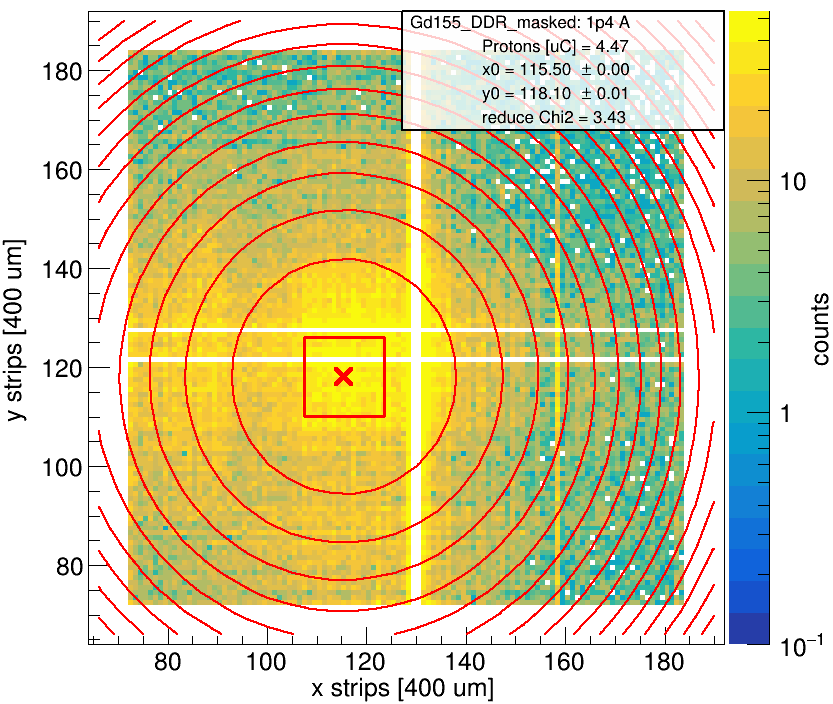}%
}
\caption{Comparison of neutron beam spots on the three different Gd samples at 0.8A and 1.4A. For the $^{157}$Gd and $^{155}$Gd samples, one can see a square region in the center, and the copper tape that was used to attach the samples to the natural Gd cathode. The red square of +/- 8 pixels around the center of the fit shows the 16 x 16 pixel zone that was considered for the intensity comparison between the different Gd isotopes. }
\label{fig: comparisonGd}
\end{figure}

The NMX instrument at ESS will use thermal and cold neutrons between 1.8 and 10 \AA. At n\_TOF, the largest neutron intensity was measured between 0.8 and 2.0 \AA~in the thermal region, as shown in Figure~\ref{fig: Thermal Peak}. For each of the different Gd samples ($^{155}$Gd and $^{157}$Gd) and the natural Gd cathode, data was taken with the beam focused on the sample. For the $^{155}$Gd and $^{157}$Gd samples, one can see a square region in the center, since copper tape was used to attach the samples to the natural Gd cathode. On the natural Gd cathode the beam profile is rounder. Using the time-of-flight method, the number of counts for wavelengths between 0.75 \AA~ and 1.85 \AA~was determined in bins of 0.1 \AA. As displayed in figure~\ref{fig: comparisonGd}, a Gaussian fit was used to find the center of the beam. A region of 16 x 16 pixels (+/- 8 pixels around the center of the fit) was defined to compare the measured intensity for different wavelengths and different Gd isotopes.

Due to the lack of an absolute efficiency reference, only the relative efficiencies of $^{157}$Gd and $^{155}$Gd compared to natural Gd can be shown. To measure the absolute detection efficiency (percentage of incoming neutrons detected), a calibrated reference detector such as a $^3$He tube would be required. In the absence of such a reference, we compare signals from different Gd isotopes measured under identical beam and detector conditions. This relative comparison, normalized to the number of incident protons, is what we refer to as relative efficiency. 

\begin{figure}[htbp]
\centering
\includegraphics[width=1.0\textwidth]{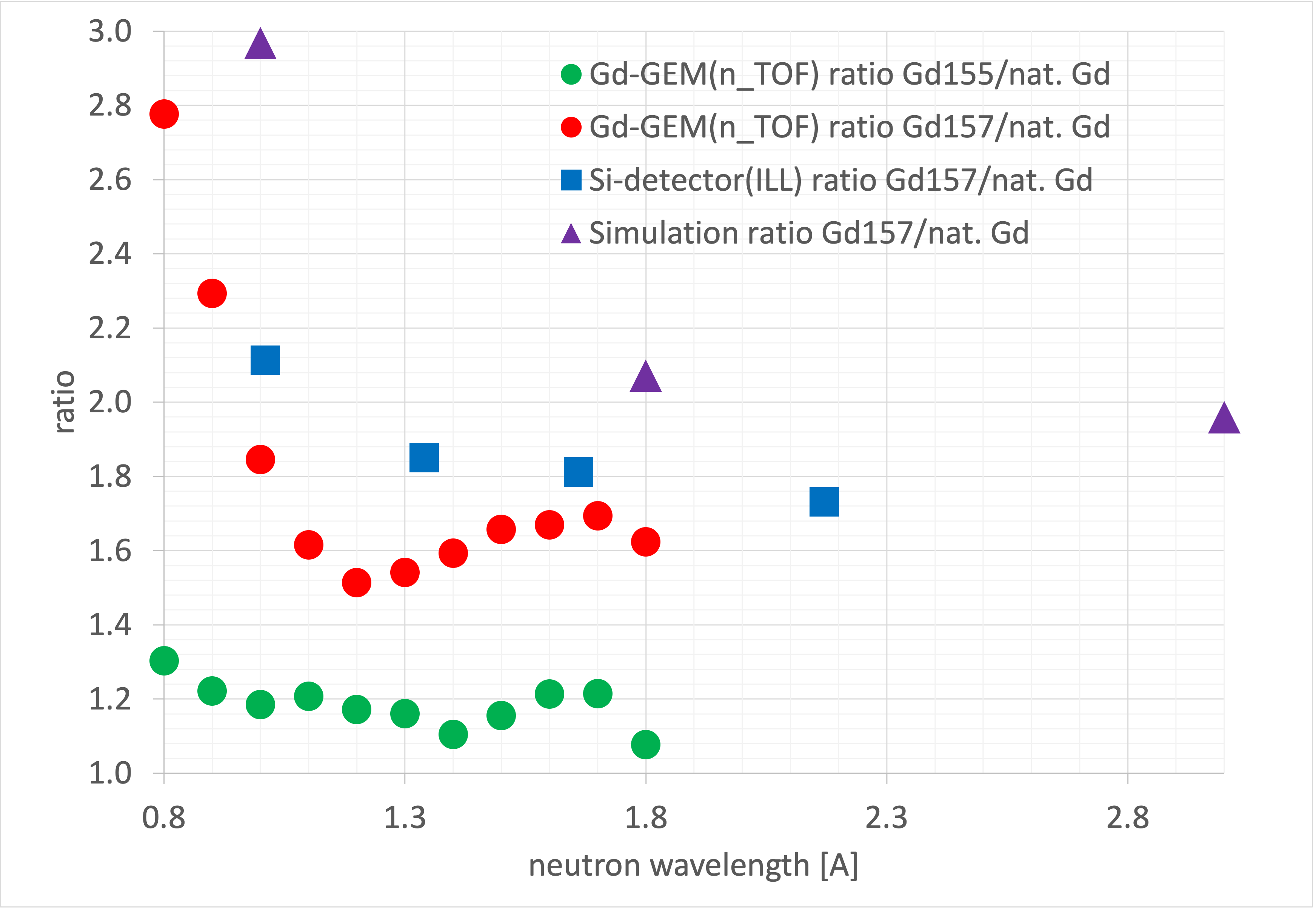}%
\caption{Relative efficiency ratios between $^{157}$Gd and natural Gd and $^{155}$Gd and natural Gd. Whereas simulations predicted for $^{157}$Gd 100$\%$ more efficiency compared to natural Gd at 1.8 A, earlier measurements at ILL with a solid state detector resulted in an 80~$\%$ improvement. With the Gd-GEM at n\_TOF about 60~$\%$ more efficiency was measured. }
\label{fig: Efficiency}
\end{figure}

Figure~\ref{fig: Efficiency} shows the n\_TOF measurements $^{155}$Gd (green) and $^{157}$Gd (red) ratios relative to natural Gd. Bruckner et. al. studied in 1999 at the ILL in Grenoble the neutron detection efficiency of Silicon detectors with a natural Gd converter and an enriched $^{157}$Gd converter in backwards configuration. The efficiency ratio that was derived from their article~\cite{Bru1999} is shown in blue. D. A. Abdushukurov published in 2013 a set of mathematical model calculations of Gd converter efficiency as a function of thickness and neutron wavelength~\cite{Abd2013}. The ratio between $^{157}$Gd and natural Gd based on his data is displayed in purple. Whereas the mathematical modelling predicts an efficiency about twice as high for $^{157}$Gd compared to natural Gd at 1.8 \AA, both experimental results show smaller gains. The solid state detector data from ILL shows an improvement of 80$\%$, and at n\_TOF with the Gd-GEM detector an improvement of 60$\%$ was measured. Bruckner et al. also observed that the absolute measured efficiency of $^{157}$Gd was higher than the calculated one, hypothesizing that the difference can be explained by the additional detection of gamma background. 

Due to the much higher neutron capture cross-section of $^{157}$Gd, naturally more neutrons are detected with $^{157}$Gd. However, the gamma sensitivity of the detector is independent of the converter material. The larger the contribution of gammas to the total count rate, the smaller the relative gain when exchanging a natural Gd converter with an enriched $^{157}$Gd converter. In addition to gamma background, also the gammas produced inside the Gd during the neutron capture can Compton scatter and produce electrons that are time-correlated with the impinging neutrons. For the particular setup with the Gd-GEM detector at n\_TOF, another explanation could be the focusing of the beam onto the enriched samples. Whereas the natural Gd cathode was large enough to see the full beam profile, the enriched samples only had an area of about 1 cm x 1 cm. The beam was not properly focused on the center of these squares, and part of the highest intensity regions were on the copper tape. For $^{157}$ Gd the highest intensity as determined by the fit is shifted from the center of the square to the right side, as visible in figure~\ref{fig: comparisonGd}. The beam was even more offset when measuring the $^{155}$Gd sample, with the highest intensity in the lower left corner of the sample. 

\begin{figure}[htbp]
\centering
\includegraphics[width=1.0\textwidth]{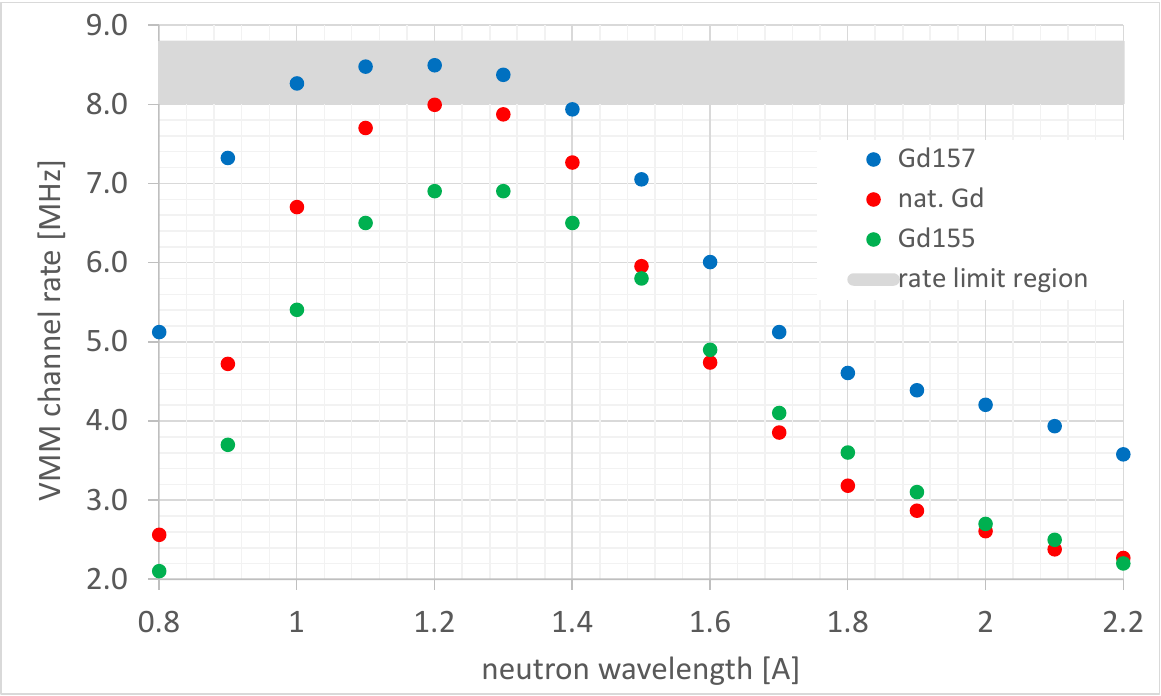}%
\caption{The theoretical absolute rate limit for one VMM3a ASIC (on the RD51 hybrid) is 8.8 MHz. At rates above 8 MHz, as reached for the VMM3a ASICs in the $^{157}$Gd region, saturation effects and thus data loss can occur. }
\label{fig: ratelimit}
\end{figure}

The highest count rates on the VMM3a ASICs were measured between 1.0 and 1.4 \AA~ in the $^{157}$Gd region. In this wavelength range the number of hits per VMM3a remains at an instantaneous rate of more than 8 million hits per second. The maximum instantaneous hit rate measured amounts to 8.6 MHz, almost reaching the absolute theoretical VMM3a hit rate limit of 8.8 MHz~\cite{Pfeiffer2022}.
As shown in~\ref{fig: ratelimit}, for natural Gd the maximum rate stays below the rate limit with a maximum of 8 MHz for 1.2~\AA~ neutrons. It is thus probable that saturation effects artificially limited the count rate during the $^{157}$Gd measurement, whereas for natural Gd the count rate stayed below the technical limits. This would explain the dip in the red curve in figure~\ref{fig: Efficiency} in this wavelength area. For the $^{155}$Gd the maximum hit rate stays well below any limit, but the measurements suffered from damaged readout strips in the center of the detector, where the sample was located. These detective strips are visible as white lines.

Taken together, gamma background and capture gamma detection, bad focusing, saturation effects and defective readout strips contributed to measuring for the enriched Gd samples a smaller relative gain increase than expected from calculations. Therefore the efficiency increase of 60$\%$ if using an enriched $^{157}$Gd converter should be seen as lower threshold of the possible efficiency gain.

\section{Conclusion}
\label{Conclusion}
The measurements presented in this paper are the first successful time-of-flight measurements with the NMX detector prototype and the ESS VMM readout. A 10 x 10~cm$^{2}$ prototype of the NMX detector was equipped with two enriched Gd samples ($^{157}$Gd and $^{155}$Gd) that were attached with copper tape to the natural Gd cathode of the detector. Three sets of measurements were taken, with the beam focused on either the natural Gd, or the $^{157}$Gd and $^{155}$Gd samples. Using the time-of-flight technique with the subsequent conversion of time-of-flight into energy, the resonant region between 1 eV and 200 eV of the $^{157}$Gd and $^{155}$Gd cross sections was studied. The peaks in the resonant region were clearly visible as having higher ADC values in the ADC spectrum. Additionally the resonant peaks had a larger number of counts per energy bin. Comparing the count rate at the center of the beam for natural Gd, $^{157}$Gd and $^{155}$Gd, the enriched $^{157}$Gd showed a efficiency that was between 180 $\%$ and 60 $\%$ larger in the thermal energy region for wavelengths between 0.8 \AA~ and 1.8 \AA. The measured 60 $\%$ increase in efficiency is smaller than expected from simulations (100 $\%$) and previous measurements with solid state detectors (80 $\%$). Detection of capture gammas and gamma background, bad focusing and saturation effects explain these deviations. The upgrade of the natural Gd converter to enriched $^{157}$Gd would thus lead to an efficiency increase of at least 60 $\%$. To decide whether the upgrade to enriched $^{157}$Gd is economically feasible, one has to study whether the savings in beam time exceed the cost of the upgrade itself.

\acknowledgments
The authors would like to acknowledge and thank the n\_TOF collaboration for lending the enriched Gd samples to ESS, and allocating beam time to us. In particular we would like to thank Alberto Mengoni, Riccardo Mucciola and Francisco Garcia Infantes for all their support during the test beams and with the interpretation of our data.

\newpage

\bibliographystyle{IEEEtran}   
\bibliography{EnrichedGd}

@article{Mastromarco2019,
	author = {M. Mastromarco and others},
	date = {2019/01/28},
	doi = {10.1140/epja/i2019-12692-7},
	id = {Mastromarco2019},
	isbn = {1434-601X},
	journal = {The European Physical Journal A},
	number = {1},
	pages = {9},
	title = {Cross section measurements of 155,157Gd(n,{\$}{$\backslash$}gamma{\$}) induced by thermal and epithermal neutrons},
	url = {https://doi.org/10.1140/epja/i2019-12692-7},
	volume = {55},
	year = {2019}
}

@article{Weiss2015,
title = {The new vertical neutron beam line at the CERN n{\_}TOF facility design and outlook on the performance},
journal = {Nuclear Instruments and Methods in Physics Research Section A: Accelerators, Spectrometers, Detectors and Associated Equipment},
volume = {799},
pages = {90-98},
year = {2015},
issn = {0168-9002},
doi = {https://doi.org/10.1016/j.nima.2015.07.027},
url = {https://www.sciencedirect.com/science/article/pii/S0168900215008566},
author = {C. Weiß and others}
}

@misc{ENDF,
 author	= {},
    year	= {},
    date = {29.01.2025},
    title	= {Evaluated Nuclear Data File (ENDF)},
    url	= {https://www-nds.iaea.org/exfor/endf.htm}
}

@misc{ESS,
    author	= {},
    year	= {},
    title	= {European Spallation Source ESS ERIC}, 
    url	= {http://europeanspallationsource.se/}
}

@misc{IFE,
    author	= {},
    year	= {},
    title = {IFE},
    url = {http://ife.no/}
}

@techreport{Peggs2013,
    author	= {S. Peggs and others},
    year	= {2013},
    title= {ESS Technical Design Report},
    institution = {European Spallation Source ERIC},
    number = {ESS-2013-0001}
}

@Article{Andersen2020,
    author = {K. H. Andersen and others},
	title = {The instrument suite of the European Spallation Source},
	year = {2020},
	journal = {Nuclear Instruments and Methods in Physics Research A},
	volume = {957},
  	pages={163402},
  	url={https://doi.org/10.1016/j.nima.2020.163402}	
}

@article{NMX2020,
    author = {M. Mark\'{o} and G. Nagy and G. Aprigliano and E. Oksanen},
    journal = {Methods in Enzymology},
    pages = {125-151},
    title = {Chapter Seven - Neutron macromolecular crystallography at the European spallation source},
    year = {2020},
    volume ={634},
    doi={https://doi.org/10.1016/bs.mie.2020.01.005}
}

@InProceedings{Kirstein2014,
    author = {O. Kirstein and others},
    journal = {Proceedings of the The 23rd International Workshop on Vertex Detectors (VERTEX2014)},
    month = {September},
    pages = {15-19},
    title = {Neutron Position Sensitive Detectors for the ESS},
    booktitle = {Proceedings of the The 23rd International Workshop on Vertex Detectors (VERTEX2014)},
    year = {2014},
    volume ={227},
    doi={https://doi.org/10.22323/1.227.0029}
}

@article{Pfeiffer2015,
  author={D. Pfeiffer and others},
  title={The uTPC method: improving the position resolution of neutron detectors based on MPGDs},
  journal={Journal of Instrumentation},
  volume={10},
  number={04},
  pages={P04004},
  url={http://stacks.iop.org/1748-0221/10/i=04/a=P04004},
  year={2015},
}

@article{Pfeiffer2016,
    author = {D. Pfeiffer and F. Resnati and J. Birch and M. Etxegarai and R. Hall-Wilton and C. H{\"{o}}glund and L. Hultman and I. Llamas-Jansa and E. Oliveri and E. Oksanen and L. Robinson and L. Ropelewski and S. Schmidt and C. Streli and P. Thuiner},
    title={First measurements with new high-resolution gadolinium-GEM neutron detectors},
    journal={Journal of Instrumentation},
    volume={11},
    number={05},
    pages={P05011},
    doi={10.1088/1748-0221/11/05/P05011},
    year={2016}
}

@Article{Sauli1997,
    author	= {F. Sauli},
    year	= {1997},
    journal = {Nuclear Instruments and Methods in Physics Research A},
    volume = {386},
    pages= {531-534},
    title = {GEM: A new concept for electron amplification in gas detectors}
}

@Article{Titov2013,
    author	= {M. Titov and L. Ropelewski},
    year	= {2013},
    journal = {Modern Physics Letters A},
    volume = {28},
    pages = {1340022},
    title = {Micro-pattern gaseous detector technologies and RD51 collaboration}
}

@Article{Guerard2012,
    author	= {B. Guerard, R. Hall-Wilton, F. Murtas},
    year	= {2013},
    journal = {CERN Document Server CDS},
    title	= {Prospects in MPGDs development for neutron detection},
    note	= {Summary based on presentations during RD51 Academia-Industry Matching Event CERN October 2013, RD51-NOTE-2014-003}, 
    url = {https://cds.cern.ch/record/2119706}
}

@Article{Altunbas2002,
    author	= {C. Altunbas},
    year	= {2002},
    journal = {Nuclear Instruments and Methods in Physics Research A},
    volume = {490},
    pages= {177--203},
    title = {Construction, test and commissioning of the triple-gem tracking detector for compass}
}

@Article{Bressan1999,
    author	= {A. Bressan and others},
    year	= {1999},
    journal = {Nuclear Instruments and Methods in Physics Research A},
    volume = {425},
    pages= {254--261},
    title = {Two-dimensional readout of GEM detectors}
}

@article{Pfeiffer2022,
title = {Rate-capability of the VMM3a front-end in the RD51 Scalable Readout System},
journal = {Nuclear Instruments and Methods in Physics Research Section A: Accelerators, Spectrometers, Detectors and Associated Equipment},
volume = {1031},
pages = {166548},
year = {2022},
issn = {0168-9002},
doi = {https://doi.org/10.1016/j.nima.2022.166548},
url = {https://www.sciencedirect.com/science/article/pii/S016890022200153X},
author = {D. Pfeiffer and others}
}

@inproceedings{DeGeronimo2012,
	author = {G. De Geronimo and others},
	title = {VMM1 - An ASIC for micropattern detectors},
	booktitle = {2012 IEEE Nuclear Science Symposium and Medical Imaging Conference Record (NSS/MIC)},
	year = {2012},
	volume = {},
	number = {},
	pages = {633--639},
	doi = {10.1109/NSSMIC.2012.6551184},
}

@article{Iakovidis2020,
	doi = {10.1088/1742-6596/1498/1/012051},
	year = {2020},
	volume = {1498},
	pages = {012051},
	author = {G. Iakovidis},
	title = {VMM3a, an ASIC for tracking detectors},
	journal = {Journal of Physics: Conference Series},
}

@article{Lupberger2018,
	author = {M. Lupberger and others},
	title = {Implementation of the VMM ASIC in the Scalable Readout System},
	journal = {Nuclear Instruments and Methods in Physics Research A},
	volume = {903},
	pages = {91--98},
	year = {2018},
	doi = {10.1016/j.nima.2018.06.046}
}

@article{Geant4a,
author = {S. Agostinelli and J. Allison and K. Amako and J. Apostolakis and H. Araujo and P. Arce and M. Asai and D. Axen and S. Banerjee and G. Barrand and F. Behner and L. Bellagamba and J. Boudreau and L. Broglia and A. Brunengo and H. Burkhardt and S. Chauvie and J. Chuma and R. Chytracek and G. Cooperman and G. Cosmo and P. Degtyarenko and A. Dell'Acqua and G. Depaola and D. Dietrich and R. Enami and A. Feliciello and C. Ferguson and H. Fesefeldt and G. Folger and F. Foppiano and A. Forti and S. Garelli and S. Giani and R. Giannitrapani and D. Gibin and J.J. G{\'o}mez Cadenas and I. Gonz{\'a}lez and G. Gracia Abril and G. Greeniaus and W. Greiner and V. Grichine and A. Grossheim and S. Guatelli and P. Gumplinger and R. Hamatsu and K. Hashimoto and H. Hasui and A. Heikkinen and A. Howard and V. Ivanchenko and A. Johnson and F.W. Jones and J. Kallenbach and N. Kanaya and M. Kawabata and Y. Kawabata and M. Kawaguti and S. Kelner and P. Kent and A. Kimura and T. Kodama and R. Kokoulin and M. Kossov and H. Kurashige and E. Lamanna and T. Lamp{\'e}n and V. Lara and V. Lefebure and F. Lei and M. Liendl and W. Lockman and F. Longo and S. Magni and M. Maire and E. Medernach and K. Minamimoto and P. Mora de Freitas and Y. Morita and K. Murakami and M. Nagamatu and R. Nartallo and P. Nieminen and T. Nishimura and K. Ohtsubo and M. Okamura and S. O'Neale and Y. Oohata and K. Paech and J. Perl and A. Pfeiffer and M.G. Pia and F. Ranjard and A. Rybin and S. Sadilov and E. Di Salvo and G. Santin and T. Sasaki and N. Savvas and Y. Sawada and S. Scherer and S. Sei and V. Sirotenko and D. Smith and N. Starkov and H. Stoecker and J. Sulkimo and M. Takahata and S. Tanaka and E. Tcherniaev and E. Safai Tehrani and M. Tropeano and P. Truscott and H. Uno and L. Urban and P. Urban and M. Verderi and A. Walkden and W. Wander and H. Weber and J.P. Wellisch and T. Wenaus and D.C. Williams and D. Wright and T. Yamada and H. Yoshida and D. Zschiesche},
journal = {Nuclear Instruments and Methods in Physics Research A},
number = {506},
pages = {250--303},
title = {Geant4 - a simulation toolkit},
doi = {10.1016/S0168-9002(03)01368-8},
volume = {506},
year = {2003}
}

@Article{Bru1999,
    author	= {G.~Bruckner et al.},
    year	= {1999},
    journal = {Nuclear Instruments and Methods in Physics Research A},
    volume = {424},
    pages= {183--189},
    title = {Position sensitive detection of thermal neutrons with solid state detectors (Gd Si planar detectors)}
}

@Article{Abd2013,
    author	= {D. A.~Abdushukurov},
    year	= {2013},
    journal = {Applied Mathematics},
    volume = {4},
    number = {8A},
    pages= {27--33},
    title = {Mathematical Modeling of the Efficiency Gadolinium Based Neutron Converters}
}

\end{document}